\documentclass[%
reprint,
showpacs,preprintnumbers,
amsmath,amssymb,
aps,
prc,
]{revtex4-2}
\pdfoutput=1
\usepackage{graphicx}
\usepackage{bm}
\usepackage{epstopdf}
\usepackage{graphicx}

\newcommand{\bb}{\mathrm{b\bar{b}}}
\newcommand{\Y}{\mathrm{\Upsilon}}
\newcommand{\YnS}[1]{{\Y({#1}\it{S})}}
\newcommand{\chib}{\mathrm{\chi_b}}
\newcommand{\chibnP}[1]{{\chib({#1}\it{P})}}

\newcommand{\RAA}[1][]{{R_\text{PbPb}^{#1}}}
\newcommand{\RAAQGP}{{\RAA[\text{QGP}]}}
\newcommand{\RAAYnS}[2][]{{\RAA[#1]\big(\YnS{#2}\big)}}

\newcommand{\Npart}{{N_\text{part}}}
\newcommand{\pT}{{p_\text{T}}}
\newcommand{\sNN}{\sqrt{s_\text{NN}}}
\newcommand{\Tcrit}{{T_\text{c}}}

\newcommand{\TDoppler}{{T_\text{D}}}
\newcommand{\tauF}{{\tau_\text{F}}}

\newcommand{\WithUnit}[2]{{#1}\,\textnormal{#2}} 
\newcommand{\To}{\textnormal{\,--\,}} 

\newcommand{\ncoll}{{\langle n_\text{coll} \rangle}}
\newcommand{\Ncoll}{{\langle N_\text{coll} \rangle}}

\newcommand{\sqrtsNN}{\sqrt{\smash[b]{s_\text{NN}}}}

\newcommand{\Tinit}{{T_0}}

\def\mus{\Upsilon{(1S)}}
\def\muss{\Upsilon{(2S)}}
\def\musss{\Upsilon{(3S)}}

\def\us{$\mus$}
\def\uss{$\muss$}
\def\usss{$\musss$}

\def\mcp{\chi_b(1P)}
\def\mcpp{\chi_b(2P)}
\def\mcppp{\chi_b(3P)}

\def\cp{$\mcp$}
\def\cpp{$\mcpp$}
\def\cppp{$\mcppp$}

\def\del{\partial}

\makeatletter
\def\@bibdataout@aps{%
\immediate\write\@bibdataout{%
@CONTROL{%
apsrev41Control%
\longbibliography@sw{%
    ,author="08",editor="1",pages="1",title="0",year="1"%
    }{%
    ,author="08",editor="1",pages="1",title="",year="1"%
    }%
  }%
}%
\if@filesw \immediate \write \@auxout {\string \citation {apsrev41Control}}\fi 
}
\makeatother

\begin{document}

\title{Bottomonium spectroscopy in the quark-gluon plasma}

\author{Georg Wolschin}
\email{wolschin@thphys.uni-heidelberg.de}
\affiliation{Institute for Theoretical Physics, Heidelberg University, Philosophenweg 16, 69120 Heidelberg, Germany, European Union}

\date{\today}

\begin{abstract}
The spectroscopic properties of heavy quarkonia are substantially different in the quark-gluon plasma (QGP) that is created in relativistic heavy-ion collisions as compared to the
vacuum situation that can be tested in pp collisions at the same center-of-mass energy. 
In this article, a series of recent works about the dissociation of the $\Upsilon(nS)$ and $\chi_b(nP)$ states in the hot QGP is summarized. Quarkonia dissociation occurs due to (1) screening of the real quark-antiquark potential, (2) collisional damping through the imaginary part of the potential, and (3) gluon-induced dissociation. In addition,  reduced feed-down plays a decisive role for the spin-triplet ground state. Transverse-momentum and centrality-dependent data are well reproduced in Pb-Pb collisions at LHC energies.
In the asymmetric p-Pb system, alterations of the parton density functions in the lead nucleus account for the leading fraction of the modifications in cold nuclear matter (CNM), but the  hot-medium effects turn out to be relevant in spite of the small initial spatial extent of the fireball, providing additional evidence for the generation of a quark-gluon droplet.
\end{abstract}

\maketitle

\section{Introduction}
\label{intro}

There is ample indirect evidence for the partonic properties of the hot and dense matter that is formed in relativistic heavy-ion collisions \cite{bjorken-1983} at energies reached at the Relativistic Heavy Ion Collider (RHIC) in Brookhaven and the Large Hadron Collider (LHC) in Geneva \cite{qm19}. Whereas probably the most convincing signals come from jet quenching in the hot medium, the modifications of quarkonia yields in heavy-ion collisions as compared to pp collisions at the same energy have turned out to be among the most sensitive signs for the properties of the hot medium:
Heavy mesons such as $J/\psi$ or $\Y$ can be used as probes for the properties of the quark-gluon plasma (QGP) that is generated in relativistic heavy-ion collisions. 

 These neutral mesons are produced in hard collisions at short formation times, typically at $\tau_{F}={0.3-0.6}${ fm/$c$}.
Among the various states that are of interest, the spin-triplet $\YnS{1}$ ground state is particularly stable. Due to its strong vacuum binding energy of about $1.10$ GeV, it has a sizeable probability to survive as a color-neutral state in the colored hot quark-gluon medium of light quarks and gluons that is generated in a central heavy-ion collision at LHC energies, even at initial medium temperatures of the order of ${400}$ {MeV} or above. The excited $\YnS{2}$, $\YnS{3}$ states with $l=0$ and the $\chi_b\text{(nP)}$ states with $l=1$ are less strongly bound, and their properties in the medium are more similar to the $J/\psi$ meson, which has a binding energy of only 0.64 GeV.

A considerable literature exists on the dissociation of quarkonia, in particular of the $\Y$~meson \cite{CMS-2012,ab14,ada14,star18}, in the hot quark-gluon medium; see Ref.\,\cite{an16} and references therein for a review. Available theoretical approaches include the works of Rapp $et~al.$ \cite{grandchamp-etal-2006,zhao-rapp-2011,em12,zhao-etal-2013,rapp17a} which is based on kinetic rate equations, Ko $et~al.$ who have investigated the temperature-dependent quarkonium formation time and bottomonia absorption in hadronic matter  \cite{ko01,ko15}, Strickland $et~al.$ with a viscous anisotropic hydrodynamical model for thermal bottomonium suppression \cite{strickland-2011,striba12,bbjs19,strickland-2019}, Zhuang $et~al.$ \cite{peng15}, and Vitev $et~al.$ \cite{vitev17,vitev19}, among others. 

It is the purpose of this article to review the theoretical approach that has been pursued in the Heidelberg Multiparticle Dynamics group
and compare the results in comprehensive form with recent data from the LHC on Pb-Pb collisions at 2.76 and 5.02 TeV, and p-Pb collisions at 8.16 TeV.

In the spatially extended fireball of gluons and light quarks that is formed in central heavy-ion collisions of symmetric systems at RHIC or LHC energies, the real part of the quarkonia potential is screened \cite{ms86} through the presence of in-medium states that differ from the vacuum state, thus reducing the yields as compared to the ones expected from pp collisions at the same energy scaled with the number of binary collisions. Hence, the nuclear modification factor $R_{AA}$ becomes smaller than one. 

This is, however, not the only modification of the quarkonia yields because the quark-antiquark potential in the medium is a complex potential with a sizeable imaginary part that accounts for collisional damping, generating a large damping width for all quarkonia states in the medium that amounts to dissociation of these states. 

The role of this complex potential is comparable to the well-known optical potentials in nuclear physics, where the imaginary part accounts for absorption into channels that are not treated explicitly. It becomes less pronounced if certain states are dealt with in a coupled-channel treatment. In quarkonia physics, one may consider gluon-induced dissociation \cite{bgw12} separately from the damping through the imaginary potential \cite{strickland-2011}, producing an additional temperature-dependent gluon dissociation width for every quarkonia state at each spacetime point in the transverse plane. 

Additional processes may be important in a proper account of quarkonia physics in relativistic heavy-ion collisions. At LHC energies, the density of charm quarks in the fireball becomes so high that their statistical recombination at freezeout generates $J/\psi$ mesons at low $p_\text{T}$  up to about 4 GeV/$c$. Hence, regeneration competes with the dissociation processes, producing a flat region in the modification factor at intermediate centralities \cite{alice20jhep} rather than a monotonic decrease towards more central collisions in 5.02 TeV Pb-Pb collisions, and probably a slight rise in central collisions at even higher LHC energies may be expected. For bottomonia at LHC energies, however, the production cross section of bottom quarks is not high enough to produce a sizeable amount in the fireball and hence, statistical regeneration plays no significant role. We shall therefore neglect it in our model, although it has been proposed \cite{rapp17a} that a detailed investigation of the $\YnS{2}$ production rates
 can help to quantify the role of regeneration from partially thermalized bottom quarks.
 
\begin{figure}[h]
\centering
\includegraphics[width=11cm]{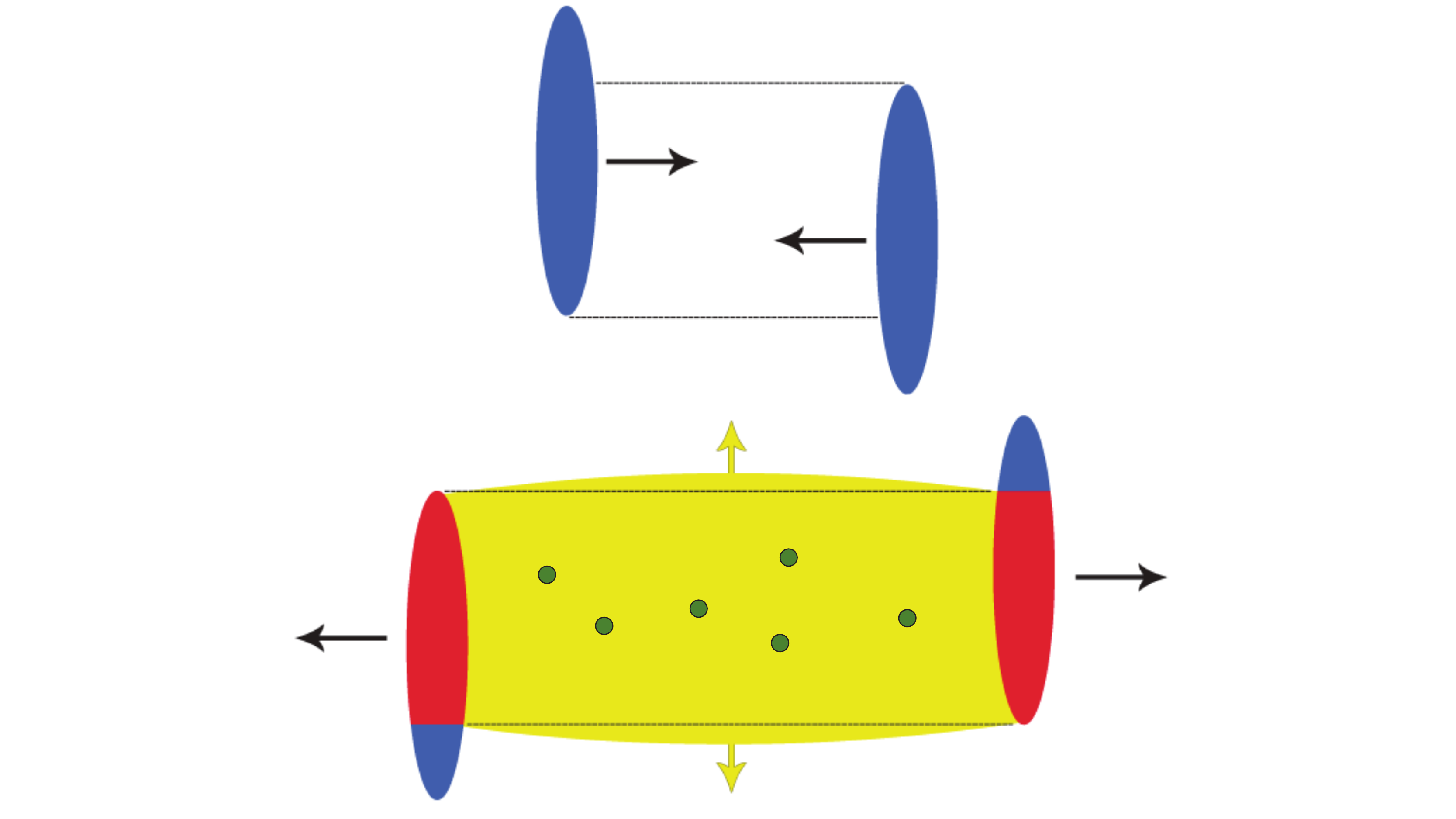}
\caption{(Color online) 
		Schematic representation of the three-source model for relativistic heavy-ion collisions at RHIC and LHC energies in the center-of-mass system: Following the collision and slowing down (\textit{stopping}) of the two Lorentz-contracted slabs (blue), the fireball region (center, yellow) expands anisotropically in longitudinal and transverse direction. At midrapidity, it represents the main source of particle production. Green circles indicate bottomonia states in the fireball. The two fragmentation sources (red) contribute to particle production, albeit mostly in the forward and backward rapidity regions. Based on a figure in Ref.\,\cite{kgw19}} 
\label{fig1}
\end{figure}

For bottomonia in heavy symmetric systems (see Fig.\,\ref{fig1}), we shall also neglect cold nuclear matter (CNM) effects, which are likely to be negligible as compared to the thermal dissociation effects in the hot fireball. Moreover, these are similar for the ground and excited states \cite{CMS-2012}, causing a renormalization of the modification factors for all $\Upsilon$ states. This would require to correspondingly adapt the initial central temperature by a small amount.
For asymmetric systems such as p-Pb, however, the CNM effects are crucial for the bottomonia modifications in the medium because most of the system remains cold, and the effects differ strongly in the forward and backward direction. We account for the corresponding modifications of the parton distribution functions (PDFs), and also coherent energy loss in 8.16 TeV p-Pb.
In symmetric and to a lesser extent also in asymmetric systems, following the dissociation in the fireball one needs to consider the reduced feed-down to lower-lying states in the medium, as compared to the situation in pp: Once the excited states are screened away or depopulated through the dissociation mechanisms, they do not contribute to the feed-down any more and correspondingly, the lower-lying states appear more suppressed as compared to pp. In particular for the $\YnS{1}$ state at LHC energies, this indirect suppression contributes a significant fraction of 30-40\% to the modification factor. We include all six states in the feed-down cascade as in Ref.\,\cite{vaccaro-etal-2013}.

In our Heidelberg model that is reviewed in this article, we thus treat the screening of the real part of the potential in the fireball, collisional damping through the imaginary part, and gluodissociation for the $l=0,1$ bottomonia states together with the reduction of the feed-down, which is significant for the $\YnS{1}$ state in symmetric systems such as Pb-Pb. We use a longitudinal boost-invariant and transverse-expanding hydrodynamical model based on a relativistic perfect fluid of the QGP  to account for the expanding background medium. For the asymmetric p-Pb system, the CNM effects dominate, but the additional dissociation in the hot quark-gluon droplet \cite{phenix19} as accounted for in our model is shown to be a relevant ingredient when comparing to LHCb \cite{lhcb18} and ALICE \cite{alice20} data. Reduced feed-down is less important in the asymmetric case.

The model ingredients are discussed in the next section. First, the treatment of screening through the real part of the quarkonia potential and damping through its imaginary part is reconsidered, with details given in the corresponding publications. Next, gluodissociation is reviewed, followed by the hydrodynamic expansion, and the feed-down cascade with its influence on the modification factors.  In Sec.\,3, our model results for symmetric systems are compared with transverse-momentum and centrality-dependent CMS Pb-Pb data at $\sqrt{s_\text{NN}}=2.76$ TeV and 5.02 TeV. In Sec.\,4, we proceed to asymmetric systems, with an emphasis on p-Pb at $\sqrt{s_\text{NN}}=8.16$ TeV, and on the importance of cold nuclear matter effects -- in particular, PDF modifications. We present the treatment of $\Upsilon$ production based on LHCb pp data, and discuss the effective path length in the medium. Hot-medium effects from the small, but expanding QGP droplet are discussed in Sec.\,5 and shown to be essential for an understanding of the LHCb and ALICE data in Sec.\,6. The conclusions are drawn in Sec.\,7.

\section{Model ingredients}
\label{ingredients}
\begin{figure*}
\centering
\includegraphics[width=14cm]{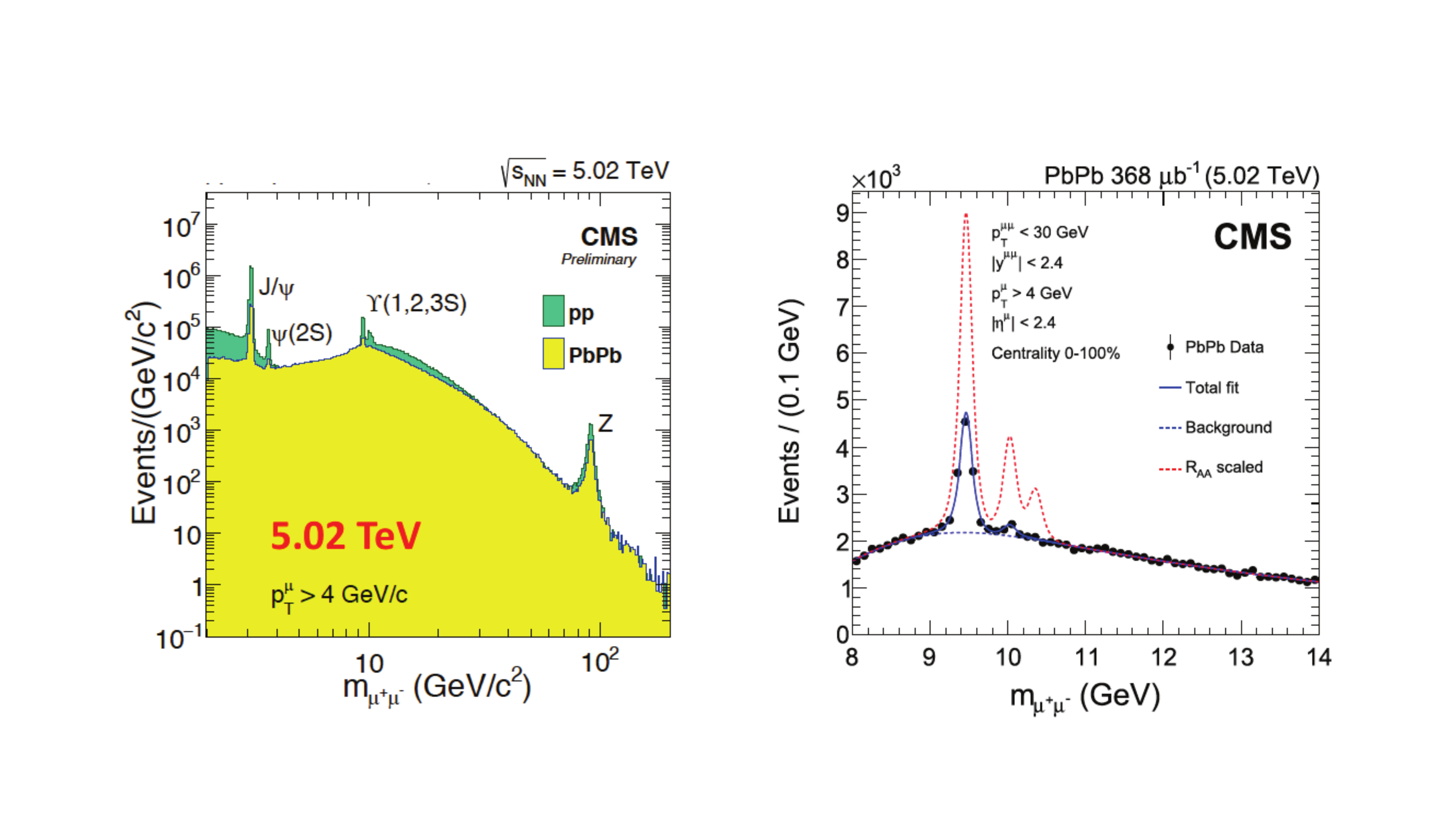}
\caption{(Color online) 
Left panel: Invariant mass distribution of muon pairs in Pb-Pb and pp collisions at $\sqrt{s_\text{NN}}=5.02$ TeV as measured by the CMS Collaboration \cite{cms19}, for the kinematic range $p_\text{T} < 30$ GeV and $y < 2.4$.
Right panel: The results of CMS fits to their data are shown as solid blue lines. To indicate the suppression of all three $\Upsilon$ states as compared to scaled pp collisions at the same energy, the amplitudes of the corresponding peaks are increased above those found in the fit by the inverse of the measured $R_\text{PbPb}$
for the corresponding $\Upsilon$ meson, dotted red lines. Reproduced with permission by Takaki (CERN) from the CMS data base, and from Ref.\,\cite{cms19}.}
\label{fig2}
\end{figure*}

The spectroscopy of charmonia \cite{seth03} and bottomonia \cite{patri13,hk20} in $e^+e^-$ and pp or p\={p} collisions  is a well-established research field. The spectra have been measured in great detail, and most of the excited spin-singlet ($\eta_c(nS)$, $\eta_b(nS))$ and spin-triplet ($J/\psi(1S), \psi(nS), \chi_c(nP); \Upsilon(nS), \chi_b(nP)$) states as well as their hadronic and other decay modes have been resolved \cite{pdg20}. The last of the states below the open-beauty threshold that has been discovered in radiative transitions by the ATLAS Collaboration in 7 TeV pp collisions at the LHC was the $\chi_b(3P)$ state \cite{ATLAS-2012}. The D0 Collaboration  subsequently confirmed the discovery based on earlier Tevatron p\={p} data taken at 1.96 TeV \cite{tevatron12}, but the decay modes still have to be determined precisely. In our work that concentrates on bottomonia, we will therefore use theoretical results from Ref.\,\cite{daghighian-silverman-1987} for the branching ratios of this particular state, whereas experimental values are taken for the other $\chi_b$ and $\Upsilon$ states.

The spacing between the $1S, 2S,$ and $3S$ levels of bottomonium resembles that in
charmonium, suggesting \cite{patri13} a simple inter-quark potential \cite{quigg79} $V (r) = C \log(r/r_0)$  that interpolates between the short-distance $\propto 1/r$ and long-distance $\propto r$ behaviors that are expected in QCD \cite{eichten80}. Many  predictions of bottomonium properties have been based on such QCD-inspired potentials, which have also been confirmed in lattice calculations, see for example Ref.\, \cite{buro17}. In this work, we will use the Cornell-type potential and its screening in the hot medium at a spacetime-dependent temperature $T$ as the real part of the diquark potential to solve the Schr\"odinger equation. In the medium, it will have to be supplemented by an imaginary part.

Bottom quarks are produced on a very short time scale of 0.02 fm/$c$ as determined from the energy-time uncertainty relation in the initial stages of the collision, before the locally thermalized plasma of light quarks and gluons is created \cite{gw20universe}. The formation time of bottomonia states is larger, in the range $\tau_\text{F}\simeq 0.3-0.6$ fm/$c$. It may differ for the individual states such as 
$\Upsilon(1S,2S,3S)$ and $\chi_b(1P,2P,3P)$, and could depend on the temperature of the emerging QGP, which would further enlarge it \cite{ko15}. The spin-triplet $\Upsilon(1S)$ state is particularly stable with a binding energy of $\simeq 1.1$ GeV, and hence, it has a sizeable probability to survive as a color-neutral state in the colored hot quark-gluon medium of light quarks and gluons that is created in a central Pb-Pb collision at LHC energies, even at initial medium temperatures of the order of 400 MeV or above. This is schematically indicated in Fig.\,\ref{fig1} that shows bottomonia states (dots) in the fireball. 

In the following, we review our model that accounts for the dissociation of the bottomonia states in the hot medium causing a suppression of these states as compared to the expectation from pp collisions at the same energy per particle pair, scaled with the number of binary collision. Since the early CMS results for the suppression of the $\Upsilon$(1S,2S) states in 2.76 TeV Pb-Pb collisions \cite{CMS-2012}, this effect has been measured very precisely both at RHIC \cite{ada14} and LHC \cite{cms17,cms19,ab14,alice19p}, in an energy region from 200 GeV to 5.02 TeV in symmetric systems (see Fig.\,\ref{fig2}). In this article, the emphasis is on the LHC results and their theoretical interpretation, as well as on predictions for the high-energy results.
\subsection{Screening and damping}
\begin{figure}[h]
\centering
\includegraphics[scale=0.22]{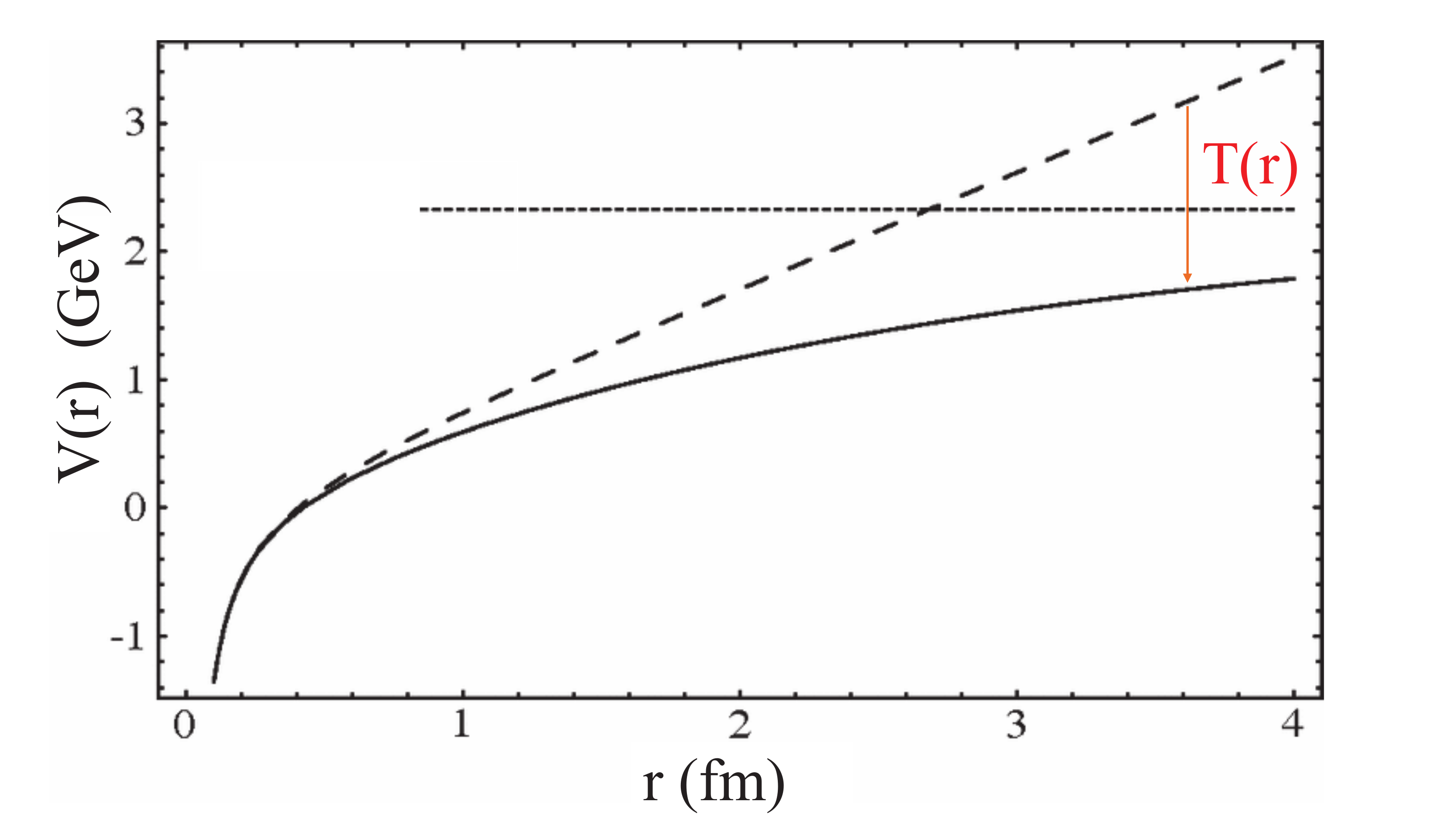}
\caption{(Color online) 
Schematic representation of the real Cornell quark-antiquark potential $V(r)$ in the vacuum (dashed curve), and in the medium at a finite temperature $T$ (solid curve). The screening effect at finite $T$ -- which is due to the presence of states different from the vacuum one -- can  cause quarkonia states to be unbound, dotted line. }
\label{fig2a}
\end{figure}

Due to the small average velocity $\sqrt{\langle v^2\rangle}{\ll}c$ of the bottom quarks in the bound states, the proper equation of motion for quarkonium states is Schr\"odinger's equation, with the color-singlet quarkonium potential. Its real part is of the Cornell type \cite{ei75} with a string and Coulomb-like part 
$V_{Q\bar{Q}} = \sigma r - \alpha_{\text{eff}}/r$,	
where $\sigma$ is the string tension \cite{ja86}, and $\alpha_{\text{eff}}$  
an effective coupling constant that accounts for the short-range gluon exchange at a given energy scale, respectively.
Since the string tension of quarkonium decreases with increasing temperature $T$ of the emerging quark-gluon medium,
the quark-antiquark binding is reduced at higher temperatures and eventually vanishes, as indicated in
Fig.\,\ref{fig2a}.
The full in-medium quark-antiquark potential also has a centrifugal term, and an imaginary part. 
We obtain the $\Upsilon$ and $\chi_b$ color-singlet wavefunctions by solving the temperature-dependent radial stationary Schr\"odinger equation.
In Refs.\,\cite{ngw14,hnw17,dhw19} we have included the effect of the running of the coupling with energy. The bound-state wave functions $\psi_{nlm}(r,\theta,\varphi,T) = g_{nl}(r,T) Y_{lm}(\theta,\varphi)/r$, with the spherical harmonics $Y_{lm}$, are characterized by the principal, angular, and magnetic quantum numbers $n$, $l$, and $m$, respectively. The $\Upsilon(nS)$ and $\chi_b(nP)$ wave functions obey the temperature-dependent radial stationary Schr\"odinger equation
\begin{eqnarray}
 \del_r^2 g_{nl}(r,T) = m_b \bigg( V_{\text{eff},nl}(r,T) - E_{nl}(T)\qquad\qquad\\\nonumber
  + \frac{i\Gamma_{\text{damp},{nl}}(T)}{2} \bigg) g_{nl}(r,T)	\label{radialschroedinger}
\end{eqnarray}
where $\Gamma_{\text{damp},{nl}}(T)$ is the damping width, $E_{nl}$ the binding energy, $m_b$ the bottom mass and $V_{\text{eff},nl}$ an effective interaction potential, that contains the centrifugal barrier and a complex interaction potential $V_{nl}$, whose real part vanishes at infinity due to screening,
\begin{equation}
 V_{\text{eff},nl}(r,T) =  V_{nl}(r,T) + \frac{l(l+1)}{m_b r^2}\,,
	\end{equation}
	\begin{eqnarray}
 V_{nl}(r,T) = - \frac{\sigma}{m_D(T)} e^{-m_D(T)r}\qquad\qquad\qquad\qquad\\\nonumber
	- C_F \alpha_{nl}(T) \Big( \frac{e^{-m_D(T)r}}{r} + iT \phi(m_D(T)r) \Big)
		  \label{complex-potential}
	\end{eqnarray}
	with the temperature-dependent Debye mass that defines the screening
	\begin{equation}
 m_D(T) = T \sqrt{4\pi \alpha_s(2\pi T) \frac{2N_c + N_f}{6}}\,,	
 \label{debyemass}
\end{equation}
and the imaginary part  \cite{laine07} that is proportional to the temperature $T$ defined through
\begin{equation}
	 \phi(x) = \int\limits_0^\infty \frac{dz \, 2z}{(1+z^2)^2} \left( 1 - \frac{\sin xz}{xz} \right)\,.
	  \label{imaginary-potential}
	 \end{equation}
	 Lattice results such as Ref.\,\cite{buro17} still deviate substantially from this form of the imaginary part.
The string tension\cite{ja86} equals $\sigma = 0.192$ GeV$^2$ and the Debye mass $m_D$ is obtained from perturbative hard thermal loop calculations. The complex potential Eq.\,(\ref{complex-potential}) is a combination of the potential found by Refs.\,\cite{laine07,brambilla-etal-2008,beraudo-etal-2008} and the non-perturbative potential ansatz of Ref.\,\cite{karsch-etal-1988}.
The number of colors and flavors, respectively, are set to $N_c = N_f = 3$ and $C_F=(N_c^2-1)/(2N_c)$. The variable $\alpha_{nl}$ denotes the strong coupling $\alpha_s$ evaluated at the soft scale $S_{nl}(T)$,
\begin{equation}
 \alpha_{nl}(T) = \alpha_s(S_{nl}(T))\,, \quad S_{nl}(T) = \langle 1/r \rangle_{nl}(T)\,.	
 \label{coupling}
\end{equation}
We have used the one-loop expression for the running of the coupling,
\begin{equation}
  \alpha_s(Q) = \frac{\alpha(\mu)}{1 + \alpha(\mu) b_0 \ln\frac{Q}{\mu}}\,,	\quad	b_0 = \frac{11 N_c - 2 N_f}{6\pi}\,,	
  \label{alpha_s}
\end{equation}
where $Q$ is the scale of four-momentum exchange and $\mu$ an arbitrary reference scale. Using values for $\alpha_s$ with matched charm- and bottom-masses \cite{bethke-2013} yields the QCD-scale $\Lambda_\text{QCD} = 276.3$ MeV.
\begin{figure}[ht]
\centering
\includegraphics[scale=0.64]{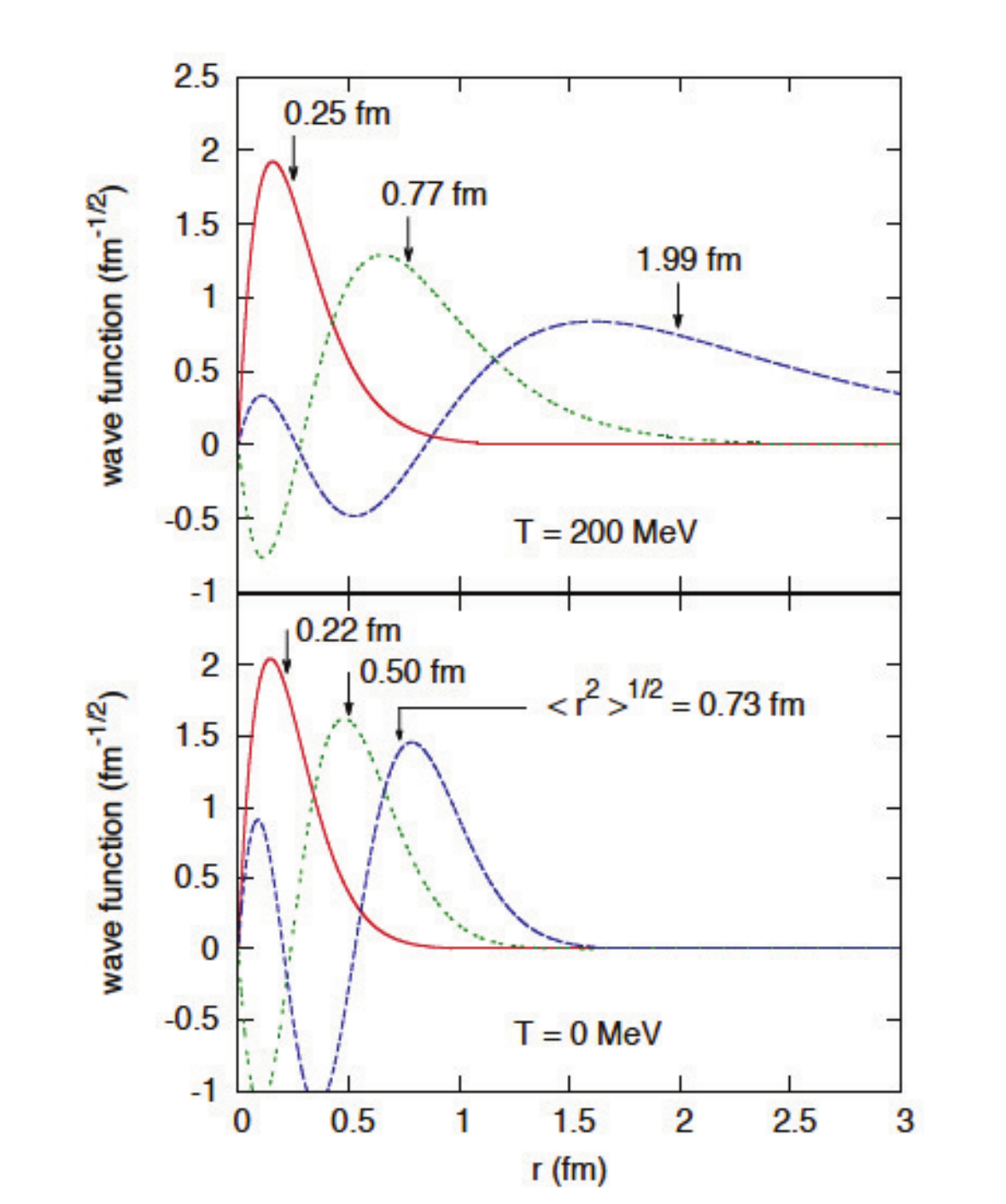}
\caption{(Color online) 
Radial wave functions of the $\YnS{1}, \YnS{2}$ and $\YnS{3}$ states (solid, dotted, dashed curves, respectively) calculated in the screened real Cornell potential for temperatures $T=0$ (bottom) and 200 MeV (top) with effective coupling constant
$\alpha_\text{eff}=0.471$, and string tension $\sigma=0.192$ GeV$^2$. The rms radii
$\langle r^2\rangle^{1/2}$ of the 2S and, in particular, 3S state strongly depend on the temperature $T$, whereas the ground state remains nearly unchanged. From Ref.\,\cite{bgw12}.}
\label{fig2b}
\end{figure}
\begin{figure*}[th]
 \centering
  \includegraphics[scale=0.64]{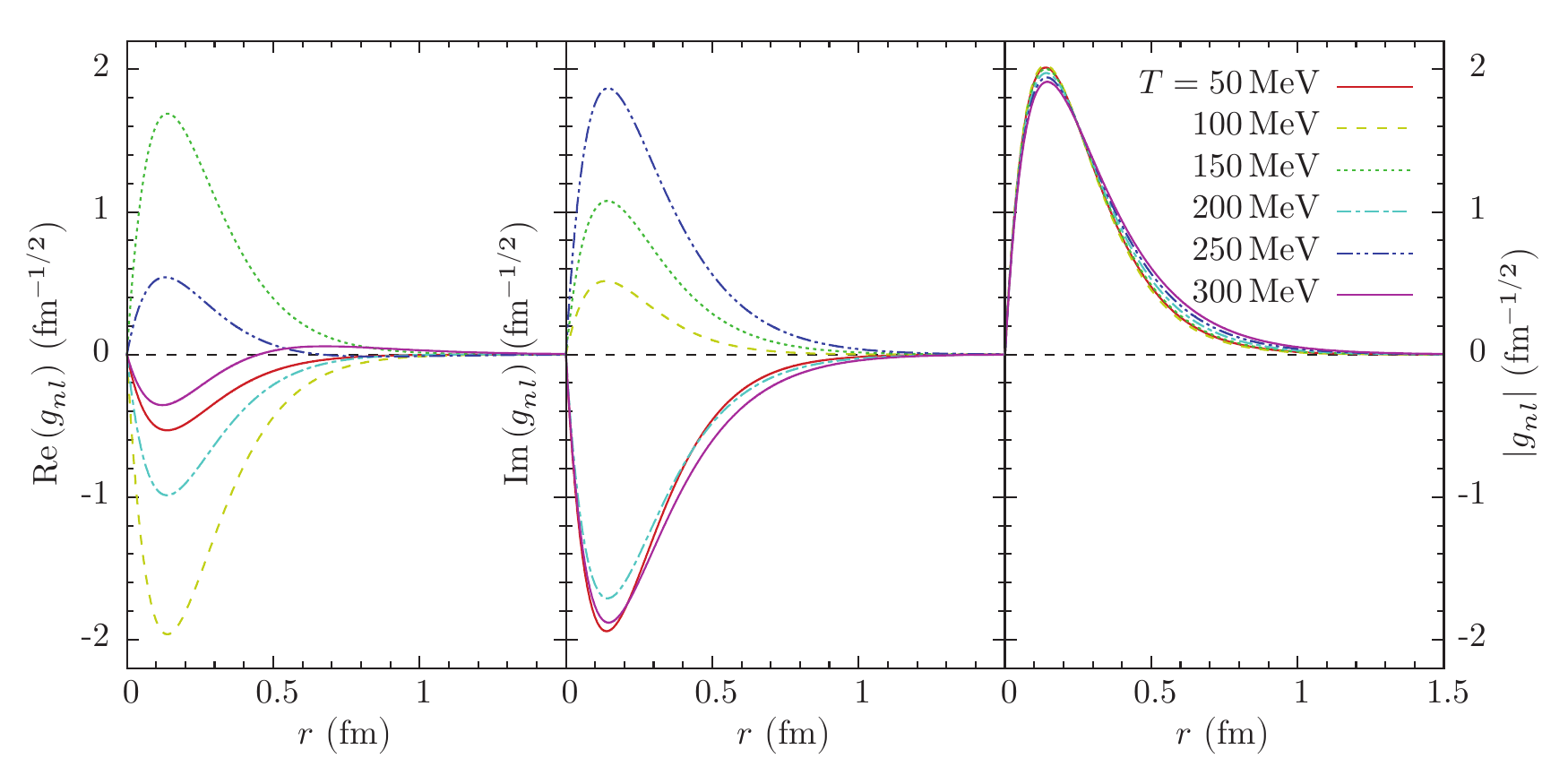}
\caption{(Color online) Radial wave functions $g_{nl}(r)$ for the $\YnS{1}$ state in the complex bottom-antibottom potential for six representative values of the temperature $T$ as obtained from  numerical solutions of Schr\"odinger's equation.  Shown are the real part (left), the imaginary part (middle), and the absolute value (right). The wave functions for $\YnS{2}$, $\YnS{3}$, and the $\chi_b$ states are calculated accordingly for each point in the transverse plane, and every temperature $T$. From Hoelck and Wolschin, unpublished.}
\label{fig3}
\end{figure*}

It is instructive to first calculate the temperature-dependent bottomonia wave functions in the real potential \cite{bgw12} for a fixed value of the effective coupling, here taken as
 $\alpha_\text{eff}=
C_F\alpha_{nl}=0.471$ . Result are shown in Fig.\,\ref{fig2b} for the $\Upsilon(1S,2S,3S)$ states at $T=0$ and $T=200$ MeV. The rms radius of the ground state is almost temperature-independent, whereas for the excited states there is a significant increase in the rms radii with rising temperature. In the full calculation with running coupling, however, it turns out that the temperature dependence of the rms radii is less pronounced due to the increasing coupling strength at small energies.

Next we have solved \cite{ngw14,hnw17} Eq.\,(1) in the complex potential numerically for the six bottomonia states $\Upsilon(nS)$ and $\chi_b(nP)~ (n=1, 2, 3)$, and different temperatures $T=0-600$ MeV. Characteristic results for the real and imaginary part of the $\Upsilon(1S)$ wavefunction plus the absolute value for six representative temperature values are shown in Fig.\,\ref{fig3}. An iterative procedure is used for the solution \cite{ngw14} since the coupling constant $\alpha_{nl}(T)$ depends on the solution of Eq.\,(\ref{radialschroedinger}): An initial value $S_{nl}^0$ is chosen in Eq.\,(\ref{coupling}) for each state and temperature, Eq.\,(\ref{radialschroedinger}) is solved in the complex $(E,\Gamma)$-plane to obtain a first approximation of the wave function, energy and damping width, then in the next step $S_{nl}^1=\langle1/r\rangle_{nl}^1$ is recalculated from $g_{nl}^1$, and used together with the energy eigenvalue and width as new initial values. As discussed in Ref.\,\cite{ngw14}, the combined effect of color screening and collisional damping
causes only a relatively small increase in the rms radii of the bottomonia states with rising $T$.

Due to the high temperature and ensuing large thermal gluon density reached in the midrapidity region, the most important single process next to screening that leads to a suppression of quarkonia at RHIC and LHC energies is gluodissociation, which is considered in the next subsection. 
\subsection{Gluodissociation}
\label{gluodissociation}
The two mechanism damping and gluodissociation emerge in different orders in the effective action, as has been shown in potential nonrelativistic~QCD approaches \cite{bram08,brambilla-etal-2011}.
The imaginary part of the interaction potential $V_{nl}$ yields collisional damping (``soft process'' in pNRQCD terminology) whereas gluodissociation is described by a singlet to octet transition (``ultrasoft process''), and hence both should be treated individually due to the separation of scales. It has not yet been investigated whether the imaginary part of the interquark potential has to be modified accordingly when gluodissociation is considered separately.

\begin{figure}[t]
 \centering
  \includegraphics[scale=0.9]{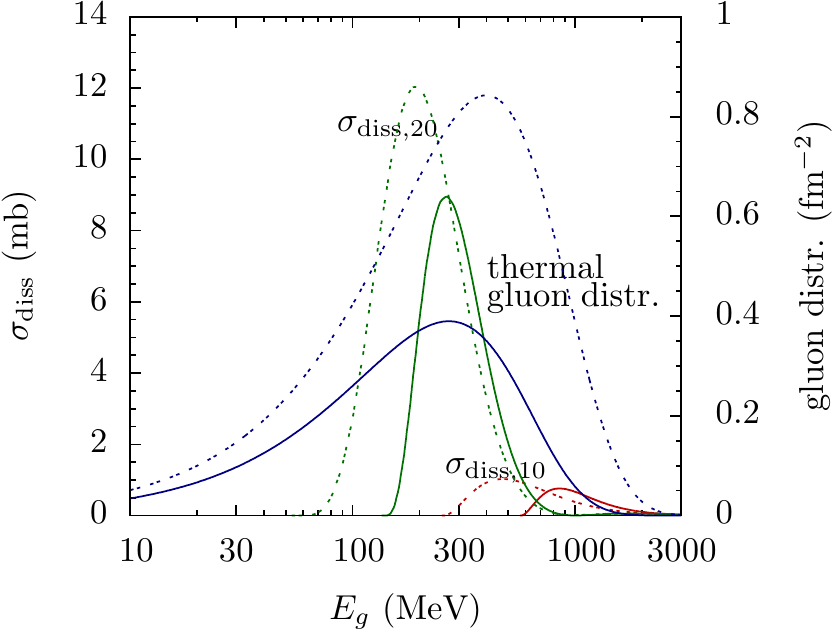}
\caption{(Color online) Gluodissociation cross section $\sigma_\text{diss}$ (left scale) of the $\YnS{1}$ (red) and $\YnS{2}$ (green) states and the thermal gluon distribution (right scale, blue curves)) plotted for temperature $T=170$ (solid curves) and $250$ MeV (dotted curves) as functions of the gluon energy $E_g$. From Nendzig and Wolschin \cite{ngw14}. \copyright~IOP Publishing.  Reproduced with permission.  All rights reserved.}
\label{fig4}
\end{figure}

The gluon energy $E_g$ has to satisfy $E_g > |E_{nl}|$ in order to be able to dissociate the bottomonium. We have calculated the gluodissociation decay width by folding the singlet-octet dipole transition cross section $\sigma_{\text{diss},nl}$ with a Bose-Einstein distribution for the gluons \cite{bgw12},
\begin{equation}
 \Gamma_{\text{diss},nl}(T) = \frac{g_d}{2\pi^2} \int\limits_0^\infty \frac{dE_g \, E_g^2 \, \sigma_{\text{diss},nl}(E_g)}{e^{E_g/T}-1}\,,	
 \label{gamma-gd}
\end{equation}
where $g_d = 16$ is the number of gluon degrees of freedom. The width $\Gamma_\text{diss}$ depends on the overlap of the gluodissociation cross section with the gluon distribution. While the peak of the Bose-Einstein distribution moves to larger gluon energies with increasing temperature the opposite is the case for the shape of the cross section (see Fig.\,\ref{fig4} for two temperature values). This results in a maximum of $\Gamma_\text{diss}$ at high temperatures for fixed coupling \cite{bgw12,ngw13}. If the running of the coupling is considered as described above, the decreasing scale for gluodissociation enhances the coupling and thus, the cross section at higher temperatures, such that the gluodissociation width keeps rising monotonically up to very high temperatures of $T>600$ MeV.

Due to the high gluon density reached at LHC energies in the mid-rapidity region, gluodissociation is a major process
besides screening and collisional damping that leads to a suppression of $\Upsilon$'s at LHC. Hence we calculate the
gluodissociation cross sections for the \us--\usss, and \cp--\cppp\ states .

The leading-order dissociation cross section of the $\bb$ states through E1 absorption of a single gluon had been derived
by Bhanot and Peskin \cite{bp79} for a Coulomb-like potential. 
We have generalized this approach in Refs.\cite{bgw12,ngw13} to include
the effect of the screened complex potential Eq.\,(\ref{complex-potential}), and obtain for a bottomonium state
$(nl)$
\begin{widetext}
\begin{eqnarray}
 \sigma_{\text{diss},nl}(E_g) &= \frac{2 \pi^2 \alpha_s^u E_g}{(2l+1) N_c^2} \sum\limits_{m=-l}^{l}
\sum\limits_{l'=0}^\infty \sum\limits_{m'=-l'}^{l'}	\nonumber
	\cdot \int\limits_0^\infty dq \, | \langle{nlm}| \, \hat{\vec{r}} \, |{ql'm'}\rangle |^2 \delta\left(
E_g+E_{nl}-\frac{q^2}{m_b} \right)	\nonumber\\
  &	= \frac{\pi^2 \alpha_s^u E_g}{N_c^2} \sqrt{\frac{m}{E_g+E_{nl}}} \, \frac{l |J^{q,l-1}_{nl}|^2 +
(l+1) |J^{q,l+1}_{nl}|^2}{2l+1}\,, 
\end{eqnarray}
\begin{eqnarray}
 J^{ql'}_{nl} &= \int\limits_0^\infty dr \, r \, g_{nl}^*(r) h_{ql'}(r)\,, 
 \label{sigma-diss}
\end{eqnarray}
\end{widetext}
with the singlet and octet states $|{nlm}\rangle$, $|{ql'm'}\rangle$ and $\alpha_s^u = \alpha_s(m_b \alpha_s^2/2) \simeq 0.59$.
The radial wave function $h_{ql'}$ of the states $|{ql'm'}\rangle$ is derived from the octet Hamiltonian with the potential
$V_8 = +\alpha_\text{eff}/(8r)$.

In the numerical calculation with running coupling, the above fixed value of $\alpha_s^u $ is then replaced by the appropriate running value.
For vanishing string tension and the corresponding values of the binding energy $E_{nl}$, a pure Coulomb $1S$
wave function, and a simplification in the octet wave function, the above expression reduces to the result in Ref.\,\cite{bp79}. Our full result for the $\Upsilon(1S)$ gluodissociation cross section agrees with the result obtained independently by Brambilla $et~al.$ using their effective field theory approach \cite{brambilla-etal-2011} in the limit discussed in
Ref.\,\cite{bgw12}.
\begin{figure}[t]
 \centering
  \includegraphics{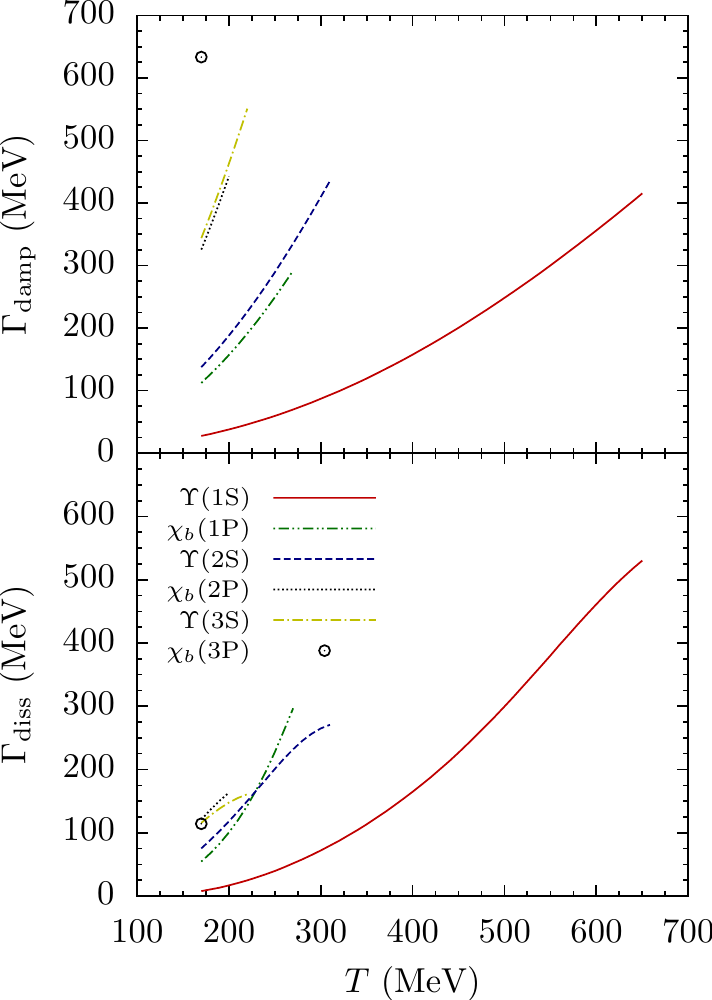}
\caption{(Color online) Partial decay widths $\Gamma_\text{damp}$ as obtained from the Schr\"odinger equation (\ref{radialschroedinger}) and $\Gamma_\text{diss}$ due to gluodissociation are plotted versus temperature $T$ for the different bottomonia. From Ref.\,\cite{ngw14}. \copyright~IOP Publishing.  Reproduced with permission.  All rights reserved.}
\label{fig4a}
\end{figure}


For each spacetime point $(t,x^1,x^2)$ in the transverse plane and for every temperature $T$ the total dissociation widths of the six screened bottomonia states are then
calculated as the incoherent sum of damping and dissociation widths,
\begin{eqnarray}
\Gamma_\text{\text{tot},{nl}}(t,x^1,x^2)=\qquad\qquad\\\nonumber
\Gamma_{\text{damp},nl}(t,x^1,x^2)+\Gamma_{\text{diss},nl}(t,x^1,x^2)\,.
\end{eqnarray}
We have taken into account dissociation through screening of the real part of the quark-antiquark potential by setting the total decay width to infinity if the energy eigenvalue of a state $|nl\rangle$ meets the continuum threshold, leading to the total hot-medium decay width \cite{hnw17,hgw17}\\
\begin{eqnarray}
	\Gamma_{\text{tot},{nl}} =\qquad\qquad\qquad\qquad\qquad\\\nonumber
	\begin{cases}
		\Gamma_{\text{damp},{nl}} + \Gamma_{\text{diss},{nl}}  \hspace{.4cm} &\text{if}\quad E_{nl} < \lim_{r\to\infty} \operatorname{Re} V_{nl}(r)\,,\\
		\infty \hspace{.4cm} &\text{else}\,.
\end{cases}
\end{eqnarray}
The temperature-dependent widths of the bottomonia states are plotted  in Fig.\,\ref{fig4a}\cite{ngw14}.
Damping and gluodissociation take place during collective expansion and cooling of the hot medium. The expansion velocity of the medium generally differs from the rms velocity of the bottomonia, which have been created with a finite transverse momentum that does not change much when they pass through the medium, due to the large bottom mass. Therefore, the relativistic Doppler effect must be taken into account \cite{esco13,hnw17} when computing the temperature-dependent dissociation widths.
In the rest frame of the bottomonia, the surrounding distribution of massless gluons acts as a Bose-Einstein distribution with an anisotropic effective temperature $\TDoppler$ that is determined from the relativistic Doppler effect as
\begin{equation}
	\TDoppler(T,|v_\text{QGP}|,\Omega) = T \frac{\sqrt{1 - |v_\text{QGP}|^2}}{1 - |v_\text{QGP}| \cos\theta}\,.
	\label{TDoppler}
\end{equation}
Here, $v_\text{QGP}$ is the average velocity of the surrounding fluid cell (measured in the bottomonium restframe) and $\Omega = (\theta,\phi)$ the solid angle where $\theta$ measures the angle between $v_\text{QGP}$ and the incident light parton.

In the rest frame of the bottomonia, the Doppler effect causes a blueshifted temperature for $\theta = 0^\circ$ and a redshifted temperature in the opposite direction $\theta = 180^\circ$.
The effects of red- and blueshift get more and more pronounced with increasing relative velocity~$|v_\text{QGP}|$, but the angular range with $\TDoppler < T$ (redshifted region) is growing while the angular range with $\TDoppler > T$ (blueshifted region) is restricted to smaller and smaller angles~$\theta$ \cite{esco13}. 

We have investigated the effect of the anisotropic temperature $\TDoppler$ on the bottomonia dissociation in Ref.\,\cite{hnw17}, where we take into account the non-existence of bound states once the effective temperature in the blueshifted region exceeds the dissociation temperature. It turns out that the anisotropic temperature results in a growing dissociation rate at larger values of transverse bottomonia momenta. This will cause a flattening of the $p_\text{T}$-dependent suppression factors $R_{AA}$ \cite{hnw17}, which would otherwise rise \cite{ngw14} at large transverse momenta. The suppression factor will be calculated in
 the next subsection, where  we take the background evolution into account through an ideal hydrodynamic description with longitudinal and transverse expansion.
\subsection{Hydrodynamic expansion}
\label{expansion}
Our treatment of the background bulk evolution with longitudinal expansion had been outlined in Ref.\,\cite{ngw13}. We have added transverse expansion -- which causes additional cooling, resulting in less pronounced dissociation --  in Ref.\,\cite{ngw14}. The following account is from Ref.\,\cite{hnw17}.

The QGP is described as a relativistic, perfect fluid consisting of gluons and massless up-, down- and strange-quarks, with an energy-momentum tensor 
\begin{equation}
 \mathcal{T} = (\varepsilon + P) u \otimes u + P,	\label{perfect-fluid-em-tensor}
\end{equation}
where $\varepsilon$ is the internal energy density of the fluid, $P$ the pressure and $u$ the fluid four-velocity.
For a general energy-momentum tensor the equations of motion are obtained by imposing four-momentum conservation, $\nabla \cdot \mathcal{T} = 0$, which yields
\begin{equation}
 \frac{1}{\sqrt{|\det g|}} \partial_\mu \left( \sqrt{|\det g|} \mathcal{T}^\mu{}_\alpha \right) = \frac{1}{2} \mathcal{T}^{\mu\nu} \partial_\alpha g_{\mu\nu}\,,
 \label{energy-momentum-conservation}
\end{equation}
where $g = g_{\mu\nu} \mathrm{d}x^\mu \mathrm{d}x^\nu$ is the spacetime-metric and Eq.~(\ref{perfect-fluid-em-tensor}) is inserted for $\mathcal{T}$. The system of equations is closed by the equation of state, appropriate for a perfect, relativistic fluid,
\begin{equation}
 P = c_\text{s}^2 \varepsilon,\qquad		c_\text{s} = \frac{1}{\sqrt{3}},\qquad		\varepsilon = \varepsilon_0 T^4.	\label{EOS}
\end{equation}
We have evaluated \cite{ngw14,hnw17} Eq.~(\ref{energy-momentum-conservation}) in the longitudinally co-moving frame (LCF), where the metric $g$ is given by
\begin{equation}
	\label{eq:metric}
	\begin{gathered}
		g = -\mathrm{d}\tau^2 + (\mathrm{d}x^1)^2 + (\mathrm{d}x^2)^2 + \tau^2 \mathrm{d}y^2,\\
		\tau = \sqrt{(x^0)^2 - (x^3)^2}\,,\qquad
		y = \operatorname{artanh}(x^3/x^0)\,.
	\end{gathered}
	\end{equation}
with the $x^1$-axis lying within and the $x^2$-axis orthogonal to the reaction plane, while the $x^3$ axis is parallel to the beam axis, and $y$ is the rapidity.
In this frame the fluid four-velocity $u$ reads
\begin{gather}\label{fluid-velocity}
 u = \gamma_\text{T} (e_\tau + v^1 e_1 + v^2 e_2)~,\\
 \gamma_\text{T} = \frac{1}{\sqrt{1 - (v^1)^2 - (v^2)^2}}~.
\end{gather}
The same transverse velocity components $v^1$, $v^2$ are measured in the laboratory frame (LF) as in the LCF. This property is convenient when dealing with quantities that depend on the transverse momentum $\pT$. Inserting Eqs.~(\ref{perfect-fluid-em-tensor}) and Eqs.\,(\ref{EOS})\To(\ref{fluid-velocity}) into Eq.~(\ref{energy-momentum-conservation}) yields
\begin{equation}
 \partial_\mu (\tau T^4 u^\mu u_\alpha) = - \frac{\tau}{4} \partial_\alpha T^4,	\qquad
 \partial_\mu (\tau \, T^3 u^\mu) = 0,	\label{EOM}
\end{equation}
where the second equation corresponds to $u \cdot (\nabla \cdot \mathcal{T}) = 0$.
\begin{figure}[tp]
 \centering
  \includegraphics{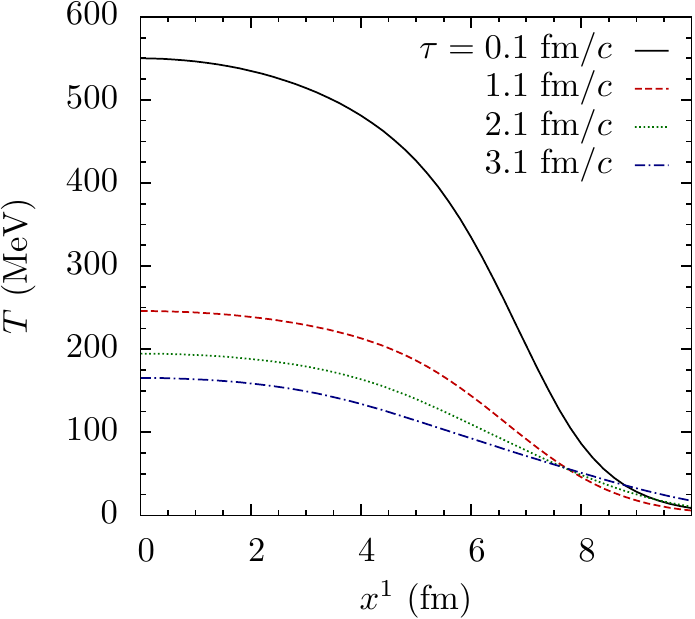}
\caption{(Color online) Profiles of temperature $T$ in the fireball plotted along the $x^1$-axis for a central collision ($b = 0$) for times $\tau = 0.1$ (solid), 1.1 (dashed), 2.1 (dotted), 3.1 fm$/c$ (dash-dotted), respectively. From Ref.\,\cite{ngw14}. \copyright~IOP Publishing.  Reproduced with permission.  All rights reserved.}
\label{fig5}
\end{figure}

We have solved \cite{ngw14,hnw17} Eqs.~(\ref{EOM}) numerically, starting at the initial time $\tau_\text{init} = \WithUnit{0.1}{fm/$c$}$ in the LCF.
The initial conditions in the transverse plane $(x^1,x^2)$ are given in Eqs.\,(14)\To(16) of Ref.\,\cite{ngw14} as
\begin{gather}
 v^1(\tau_{\text{init}}) = v^2(\tau_{\text{init}}) = 0 \\
 T(b, \tau_{\text{init}}, x^1, x^2) = T_0 \left( \frac{N_\text{mix}(b,x^1,x^2)}{N_\text{mix}(0,0,0)} \right)^{1/3} \\
 N_{\text{mix}}^{\text{RHIC}} = \frac{1 - f}{2} N_{\text{part}} + f N_{\text{coll}},\quad f = 0.145 \\
 N_{\text{mix}}^{\text{LHC}} = \hat{f} N_{\text{part}} + (1-\hat{f}) N_{\text{coll}},\quad \hat{f} = 0.8
\end{gather}
where $f,\hat{f}$ are taken from Refs.\,\cite{PHOBOS-2004,ALICE-2011a} and $b$ is the impact parameter.
The initial central temperature $T_0$ is fixed through a fit of the $\pT$-dependent minimum-bias experimental $\RAAYnS{1}$ results for Pb-Pb at $\WithUnit{2.76}{\,TeV}$.
Corresponding profiles of temperature $T$ in the fireball plotted along the $x^1$-axis for a central collision at different times are shown in Fig.\,\ref{fig5} from Ref.\,\cite{ngw14}.

Here, the initial central temperature $T_0$ is somewhat higher than in Ref.\,\cite{hnw17}, where we have incorporated the relativistic Doppler effect. It requires a lower initial central temperature, because the effective temperature $T_\text{eff}$ that the moving bottomonia see in the medium is enhanced when the Doppler effect is considered.
For other systems and incident energies, $T_0$ is scaled consistently with respect to the produced charged hadrons, thus providing predictions for the bottomonia suppression in
5.02 TeV Pb-Pb, for example.

It is recognized that there are more sophisticated hydrocodes available -- in particular, those including viscosity. Cooling proceeds slower in a viscous medium and hence,
the quarkonia states will be more suppressed because the QGP phase last longer. Such a more detailed treatment of the background evolution would then require a rescaling of the two parameters that enter our model, namely, the bottomonia formation time, and the initial central temperature $T_0$.

We have defined the QGP-suppression factors $R^\text{QGP}_{AA,nl}(c,\pT)$ which quantify the amount of in-medium suppression of bottomonia with transverse momentum $\pT$ for Pb-Pb collisions in the centrality bin $c$, where $b_c \leq b < b_{c+1}$. The QGP-suppression factor is not directly measurable since it accounts only for the amount of suppression inside the fireball due to the three processes color screening, collisional damping and gluodissociation. It is given by the ratio of the number of bottomonia that have survived the fireball to the number of bottomonia produced in the collision. The latter scales with the number of binary collisions at a given point in the transverse plane and hence with the nuclear overlap, $N_\bb \propto N_\text{coll} \propto T_{AA}$. Thus we define $R_{AA,nl}^\text{QGP}$ as 
\begin{eqnarray}
\label{RAAQGP}
 R_{AA,nl}^\text{QGP}(c,\pT) =\qquad\qquad\qquad\qquad\\\nonumber
 \frac{\int_{b_c}^{b_{c+1}} \mathrm{d}b \, b \int \mathrm{d}^2x \, T_{AA}(b,x^1,x^2) \, D_{nl}(b,\pT,x^1,x^2)}{\int_{b_c}^{b_{c+1}} \mathrm{d}b \, b \int \mathrm{d}^2x \, T_{AA}(b,x^1,x^2)}\,.
\end{eqnarray}
The damping factor $D_{nl}$ is determined by the temporal integral over the corresponding $\bb$ decay width $\Gamma_{nl}$,
\begin{eqnarray}
\label{damping-factor}
 D_{nl}(b,\pT,x^1,x^2)=\qquad\qquad\qquad\\\nonumber
 \exp\left[ - \int\limits_{\tau_{\text{F},nl} \gamma_{\text{T},nl}(\pT)}^\infty \frac{\mathrm{d}\tau \, \Gamma_{nl}(b,\pT,\tau,x^1,x^2)}{\gamma_{\text{T},nl}(\pT)}\right],
\end{eqnarray}
where $\tau_{\text{F},nl}$ is the formation time in the bottomonium rest-frame, $\gamma_{\text{T},nl}(\pT) = \sqrt{1 + (\pT/M^{\text{vac}}_{nl})^2}$ the Lorentz-factor due to transverse motion in the LCF, and $M^{\text{vac}}_{nl}$ the experimentally measured bottomonium vacuum mass.
\begin{table}[h]
\caption{Melting temperatures $T_\text{m}$ of the different bottomium states. No bound state solutions to Eq.\,(\ref{radialschroedinger}) exist for $T > T_\text{m}$.}
\label{tab1}
 \vspace{3mm}
 \centering
\begin{tabular}{cr@{\hspace{2em}}}	\hline\hline
  State	&	\multicolumn{1}{c}{$T_\text{m}$ (MeV)}	\\	\hline
  \us   & 655	\\
  \cp   & 273	\\
  \uss   & 320	\\
  \cpp   & 206	\\
  \usss   & 228	\\
  \cppp   & $\sim 175$	\\	\hline\hline
\end{tabular}
\end{table}
\begin{figure*}[th]
 \centering
  \includegraphics[scale=0.32]{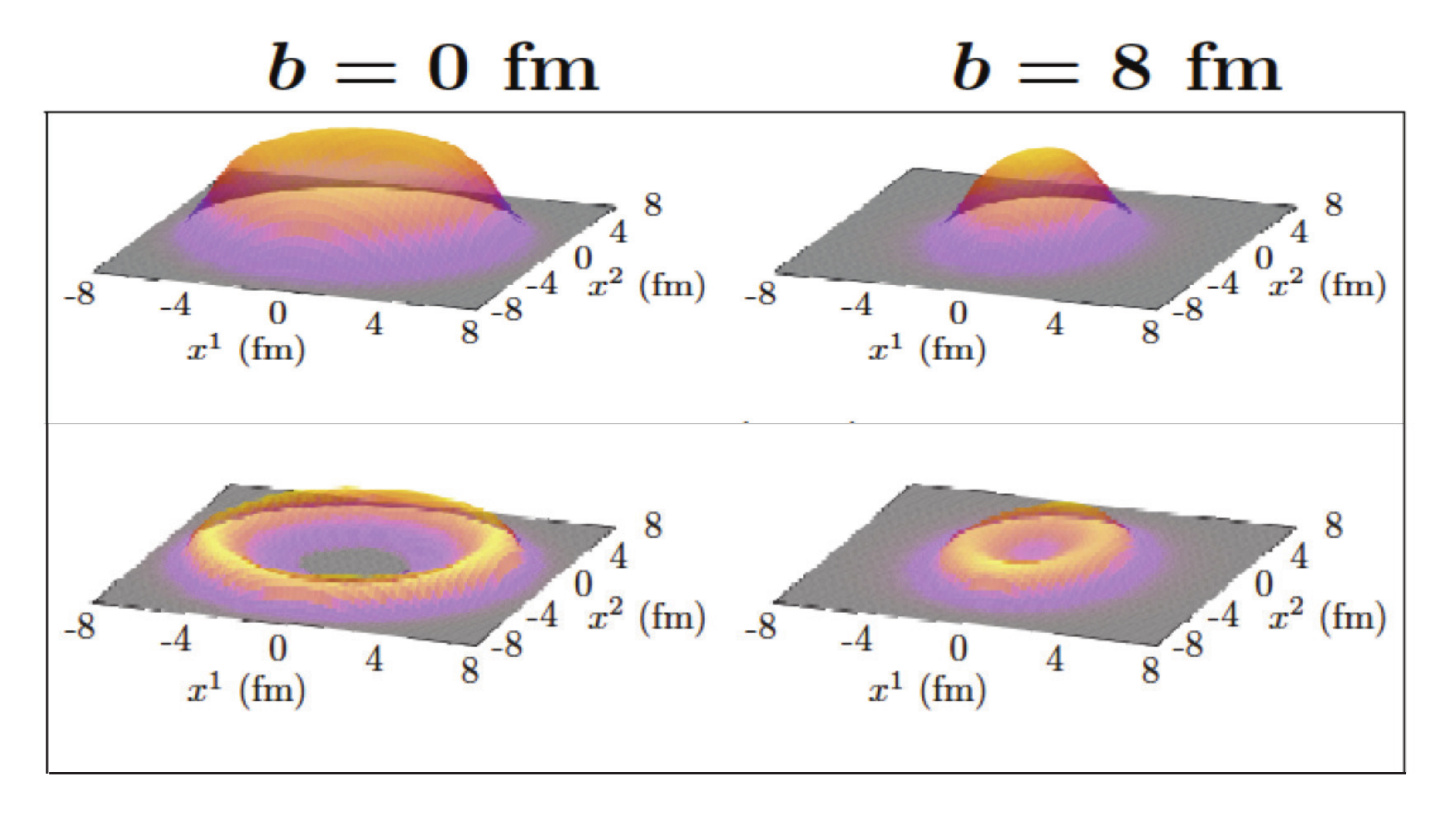}
  \caption{(Color online) Two-dimensional profiles of the integrand in Eq.\,(\ref{RAAQGP}), $T_{AA} D_{n0}$, for the $\YnS{1}$ (top) and $\YnS{2}$ (bottom) states in central ($b = 0$, left) and peripheral ($b = 8$ fm, right) 2.76 TeV Pb-Pb collisions with $p_\text{T} = 0$. The profiles scale with the fraction of $\YnS{1}$ and $\YnS{2}$ that survive dissociation throughout the lifetime of the QGP. The suppression is more pronounced for the excited state, and at smaller impact parameters. Based on Ref.\,\cite{ngw14}. \copyright~IOP Publishing.  Reproduced with permission.  All rights reserved.}
\label{fig6}
\end{figure*}
The weighted damping factor $T_{AA} D_{nl}$  scales with the number of surviving bottomonia in the transverse plane $(x^1,x^2)$.  Two-dimensional plots of $T_{AA} D_{n0}$ are shown in Fig.\,\ref{fig6} for 2.76 TeV Pb-Pb, where in this case a formation time $\tau_{F,nl} = 0.5$ fm/$c$ has been used \cite{ngw14}. The transverse $T_{AA} D_{n0}$-distributions of the \us\ and \uss\ states are displayed for a central ($b = 0$) and a peripheral collision ($b = 8$ fm) with $p_\text{T} = 0$. Most bottomonia are formed in the center of the heavy-ion collision, where many binary nucleon-nucleon collisions occur. Due to the high central temperatures, however, strong suppression changes the shape of the surface for \us\ from cone-like (peripheral) to volcano-like (central). The first excited state \uss\ is suppressed much more efficiently than the more stable ground state \us. The action of color screening forbids the formation of bound  states above the melting temperature $T_\text{m}$ (Table\,\ref{tab1}) and thus can enforce $D_{nl} = 0$ in close proximity to the collision center. This does not occur, however, for the \us\ if its melting temperature (see Table\,\ref{tab1}) is higher than the maximum fireball temperature, $T_\text{m} > T_0$. Here, the melting temperatures $T_\text{m}$ are calculated from the condition that no bound-state solution of Eq.\,(1) with the complex potential Eq.\,(2) exists for $T>T_\text{m} $.
\subsection{Initial bottomonia populations}
To estimate the initial populations $N^\text{i}_{AA,nl}$ of the six bottomonia states that we have treated explicitly, we consider the measured final populations $N^\text{f}_{\text{pp},nl}$ of the three $\YnS{n}$-states in pp collisions at the same energy and calculate the decay cascade~\cite{vaccaro-etal-2013} backward to obtain the initial populations in pp, $N^\text{i}_{\text{pp},nl}$, shown in Table~\ref{tab2}.
These are then scaled by the number of binary collisions $N_\text{coll}$ yielding the initial populations $N^\text{i}_{AA,nl}$ in the heavy-ion case. 
When the suppression factors are calculated, the number of binary collisions cancels out by definition.
The required branching ratios are taken from the Review of Particle Physics~\cite{pdg20} or from theory where no experimental values are available (as is the case for 
$\chibnP{3}$), see~\cite{vaccaro-etal-2013} for details and references.
\begin{table}
	\caption{\label{tab2}
		Initial populations of the different bottomonium states as obtained from an inverted feed-down cascade calculation in pp collisions at $\WithUnit{2.76}{TeV}$ normalized by the $\YnS{1}$ population after feed-down, $n^\text{i}_{\text{pp},nl} = N^\text{i}_{\text{pp},nl}/N^\text{f}_{\text{pp},\YnS{1}}$.
		($\Gamma_\chibnP{3}$ denotes the yet unkown vacuum decay width of the $\chibnP{3}$ state which cancels out in the computation of final populations.)
	}
	\bigskip
	\centering
	\begin{tabular}{rr}
		\hline\hline
		State & $n^\text{i}_{\text{pp},nl}$ \\
		\hline
		$\YnS{1}$ & $0.373$ \\
		$\chibnP{1}$ & $1.084$ \\
		$\YnS{2}$ & $0.367$ \\
		$\chibnP{2}$ & $0.881$ \\
		$\YnS{3}$ & $0.324$ \\
		$\chibnP{3}$ & $0.00835\,\text{eV}^{-1} \Gamma_\chibnP{3}$ \\
		\hline\hline
	\end{tabular}
\end{table}

Regarding the production process, we take the same formation time of $\WithUnit{0.4}{fm/$c$}$ for ground and exited states, with step functions for the production as function of proper time.
In the co-moving coordinate system (LCF) used for the hydrodynamical calculation, time dilation of the formation times is then taken into account.
As has been indicated e.g. in Ref.\, \cite{ko15}, the quarkonium formation time in heavy-ion collisions is not well determined. We have investigated the dependence of our model results on the formation time to some extent in Ref.\,\cite{ngw14}, but have kept it fixed  for all states in subsequent work \cite{hnw17}. Clearly this is a simplification, and the medium and temperature effects on the formation time need further investigation \cite{ko15,gami15}.
\begin{figure}
\centering
\includegraphics[scale=0.46]{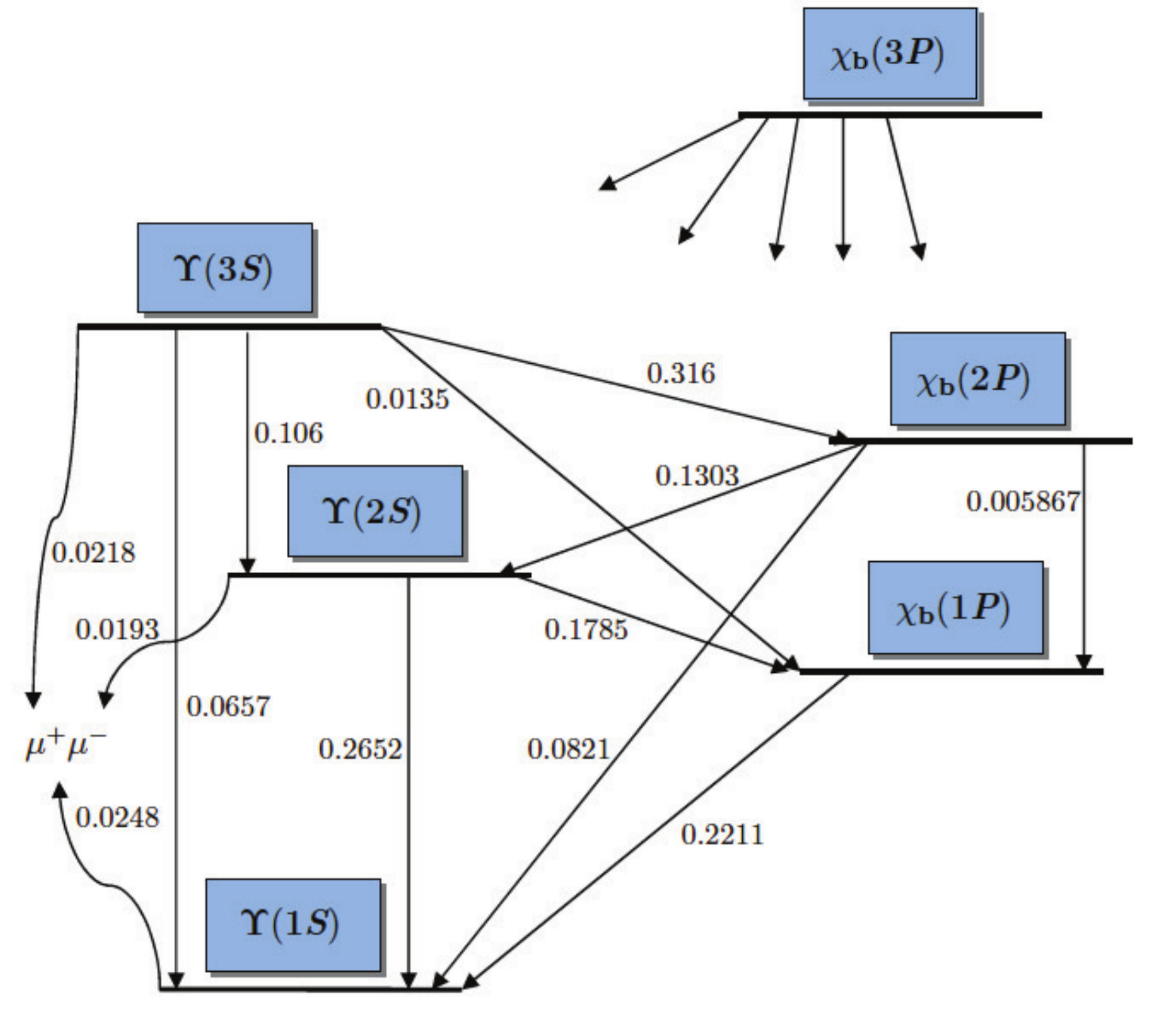}
  \caption{(Color online) Branching ratios for decays within the bottomonium family and into $\mu^\pm$-pairs according to the particle data group. Branching ratios for the  $\chi_b(3P)$ state are still unknown, partial
decay widths are taken from theory \cite{daghighian-silverman-1987}. From Ref.\,\cite{vaccaro-etal-2013}.}
\label{fig7}
\end{figure}

Once the bottomonia states have survived the dissociation processes in the hot quark-gluon plasma environment, the feed-down cascade from the excited states to the ground state is considered in detail \cite{vaccaro-etal-2013}. The corresponding branching ratios as taken from the particle data group \cite{pdg20}, and from theory \cite{daghighian-silverman-1987}
for the $\chi_b(3P)$ state, are shown in Fig.\,\ref{fig7}.
Due to the rapid melting or depopulation of the excited states caused by the mechanisms in the QGP-phase, the feed-down to the ground state is reduced, resulting in additional 
$\YnS{1}$-suppression with respect to the situation in pp~collisions at the same energy. For the excited states, in contrast, the reduction of the feed-down does not substantially modify the suppression factors.

The relevance of reduced feed-down for a given bottomonium state as function of incident energy (RHIC vs. LHC), and of its relative importance for the $\YnS{1}$ and $\YnS{2}$ states, has been a focal point in Ref.\,\cite{hnw17}. As will be shown in the following section, indeed a substantial fraction
of the suppression for the $\YnS{1}$ state is found to be due to reduced feed-down, and -- as detailed in Ref.\,\cite{hnw17} -- even more so in U-U collisions at the RHIC energy of 193 GeV.
In contrast, most of the $\YnS{2}$ suppression is caused by the hot-medium effects.
\subsection{Anisotropic bottomonium flow?}
It is of interest to determine if the bottomonia distributions in more peripheral collisions become anisotropic, as has been found for produced particles in general. The quadrupole part of the momentum anisotropy is caused by the almond-shaped spatial anisotropy of the overlap region, which translates to momentum space. It is more pronounced for lighter mesons
such as pions, and can be quantified by the ellipticity $v_2$ of the momentum distribution in a Fourier decomposition of the experimentally determined, event-averaged particle distribution \cite{hesne13} 

\begin{equation}
\frac{d\langle N \rangle}{d\phi}=\frac{\langle N \rangle}{2\pi}\left(1+2\sum_{n=1}^{\infty}\langle v_n \rangle \cos[n(\phi-\langle \psi_n \rangle)]\right)
\end{equation}
with the azimuthal angle $\phi$, the mean flow angle $\langle \psi_n \rangle$,  and $\langle N \rangle$ 
the mean number of particles of interest per event (charged hadrons or
identified particles of a specific species). Since the flow planes are not experimentally known, the anisotropic flow coefficients are
obtained using azimuthal angular correlations between the observed particles. The experimentally reported anisotropic flow coefficients from two-particle correlations can then be obtained as the
root-mean-square values, $v_n\{2\} \equiv v_n\equiv\sqrt{\langle v_n^2 \rangle} $, and the flow coefficients are being measured not in individual events, but in centrality classes.

Whereas for charged hadrons the flow coefficients have been measured very precisely \cite{hesne13} with $v_2$-values up to $20-30\%$ for pions, kaons, and antiprotons and maxima near $p_\text{T}\simeq 3$ GeV/$c$, this is more difficult for quarkonia due to the much smaller production rates. For the charmonium spin-triplet ground state, the ellipticity coefficient at  forward rapidity ($2.5<y<4$) in the centrality class $5-60\%$ is $v_2\simeq3-8\%$, depending on $p_\text{T}$ \cite{alice19a}, implying that $J/\psi$ shows elliptic flow, albeit on a smaller scale due to the larger mass. 

It may, therefore, appear possible that even bottomonium exhibits flow, following the mass ordering of lighter particles resulting from a collective
expansion of the medium \cite{rey20}. Indeed, the large statistical error bars on the presently available bottomonium data \cite{alice19a} for $v_2$ do not yet exclude such a possibility,
although it is quite doubtful whether this meson -- which is about three times heavier than charmonium -- flows with the expanding hot medium. Instead, the $\Upsilon(1S)$ is expected to essentially maintain its trajectory in the hot QGP, unless it is dissociated. Still, its momentum distribution in more peripheral collisions may exhibit a finite $v_2$ due to the anisotropic escape from the fireball, because the path length from $\Upsilon$ formation to escape in the transverse plane depends on the azimuthal angle in more peripheral collisions. 

This mechanism had already been suggested for $J/\psi$ by Wang and Yuan at SPS and RHIC energies \cite{wang02}, and has been used in Ref.\,\cite{bbjs19} for the bottomonium states in non-central 2.76 TeV Pb-Pb. There, the maximum of $v_2[\Upsilon(1S)]$ including feed-down contributions is found to be below $1\%$ in the $40-50\%$ centrality class.
The anisotropy is very small, compatible with zero. Although we have performed a corresponding calculation of $v_2$ in the $5-60\%$ centrality class that is consistent with the results of Ref.\,\cite{bbjs19} for the $p_\text{T}$ dependence, it is too early to present a comparison with the available data: For more definite conclusions, one has to wait for a reduction of the experimental uncertainties in run 3.
\section{Model results for symmetric systems}
\begin{figure}
\centering
\includegraphics[scale=0.7]{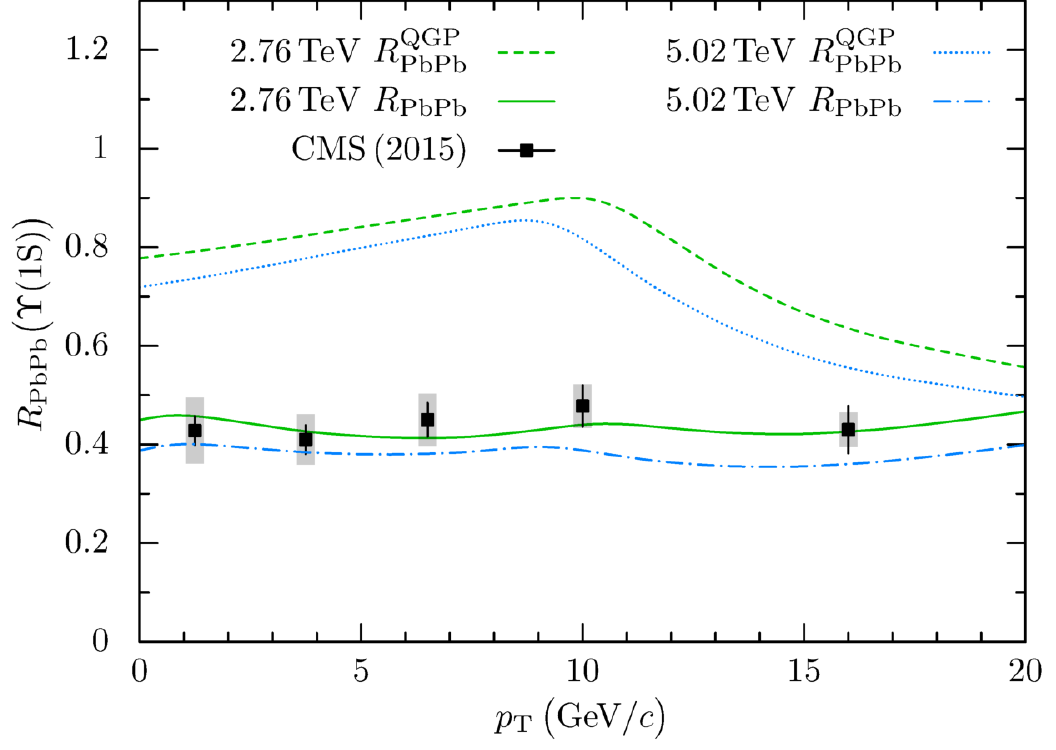}
\caption{(Color online) 
Transverse-momentum dependence of the suppression factors $R_\text{PbPb}^\text{QGP}$ (dashed line) and $R_\text{PbPb}$ (solid line) for the ground state in minimum-bias Pb-Pb~collisions at $\sqrt{s_\text{NN}}=2.76$ TeV ($T_0 = 480$ {MeV}) compared with CMS data~\cite{cms17}.
		Our predictions for $\sqrt{s_\text{NN}}=5.02$ TeV ($T_0 = 513$ {MeV}) are shown as dotted and dash-dotted curves for $R_\text{PbPb}^\text{QGP}$ and $R_\text{PbPb}$, respectively. From {Ref.\,\cite{hnw17}.}}
\label{fig8}
\end{figure}

\begin{figure}
\centering
\includegraphics[scale=0.7]{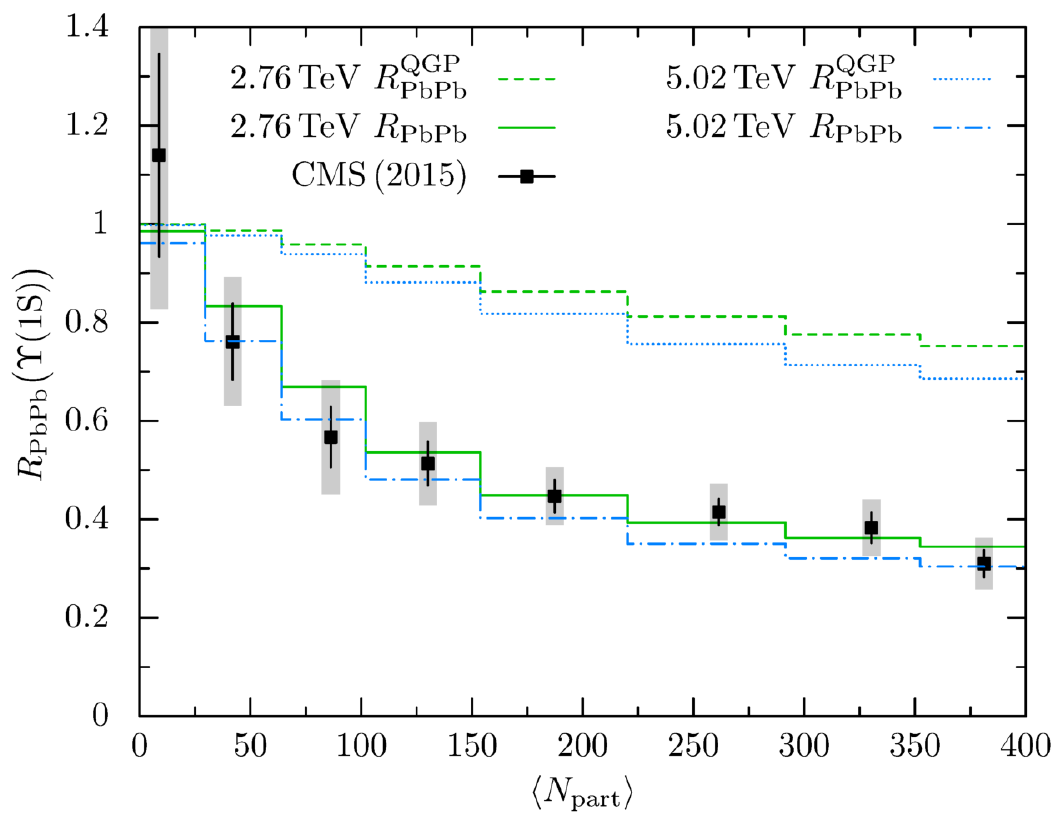}
\caption{(Color online) Calculated suppression factor $\RAAYnS{1}$ in Pb-Pb~collisions at $\sNN = \WithUnit{2.76}{TeV}$ (lower solid line) and prediction at $\WithUnit{5.02}{TeV}$ (lower dash-dotted line) together with centrality-dependent $\WithUnit{2.76}{TeV}$ data from CMS (squares, $|y| < 2.4$,~\cite{cms17}) as function of the number of participants $\Npart$ (averaged over centrality bins).
		The suppression factors $\RAAQGP$ in the QGP-phase without the effect of reduced feed-down are shown as upper curves (dashed and dotted) for both energies, again yielding slightly more suppression at the higher energy. From Ref.\,\cite{hnw17}.}
\label{fig9}
\end{figure}
\begin{figure}
\centering
\includegraphics[scale=0.74]{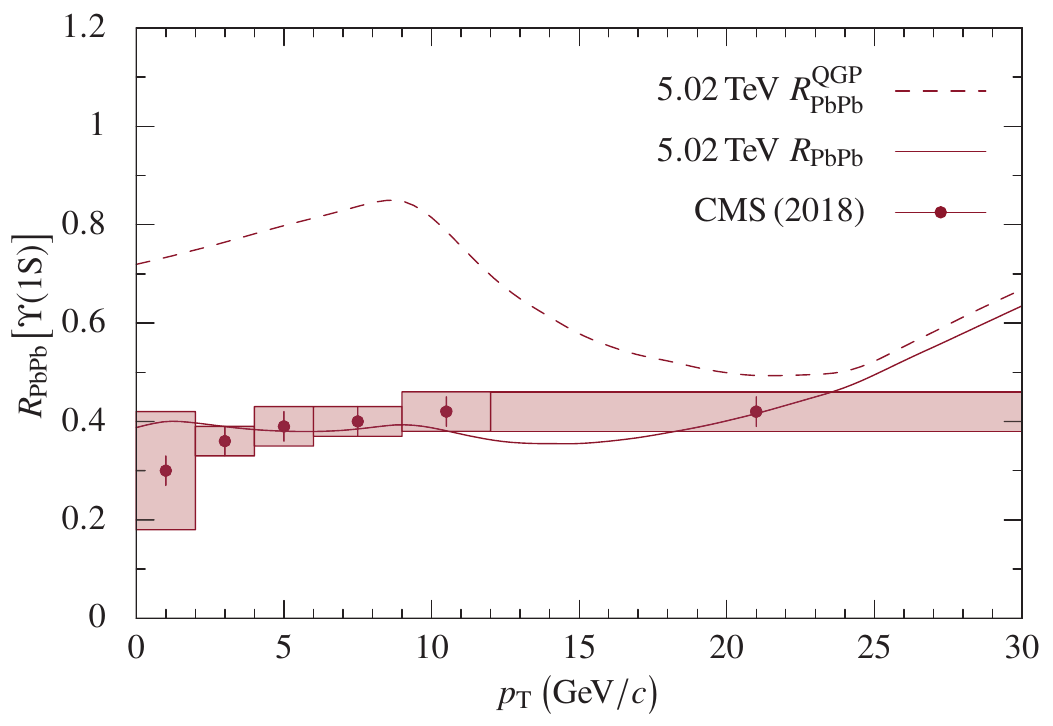}
\caption{(Color online) Transverse-momentum dependence of the modification factor $R^\text{QGP}_\text{PbPb}[\Upsilon(1S)]$ in the hot medium (dashed curve), and of the total suppression 
$R_\text{PbPb}[\Upsilon(1S)]$ including reduced feed-down (solid curve) for the $\Upsilon(1S)$ state in minimum-bias Pb-Pb~collisions at $\sNN = {5.02}$ {TeV} ($T_0 = {513}$ {MeV}). Recent CMS data from Ref.\,\cite{cms19} are compared with our theoretical prediction (Fig.\,12, above) from Ref.\,\cite{hnw17}.}
\label{fig10}
\end{figure}
\begin{figure}
	\centering
\includegraphics[scale=0.7]{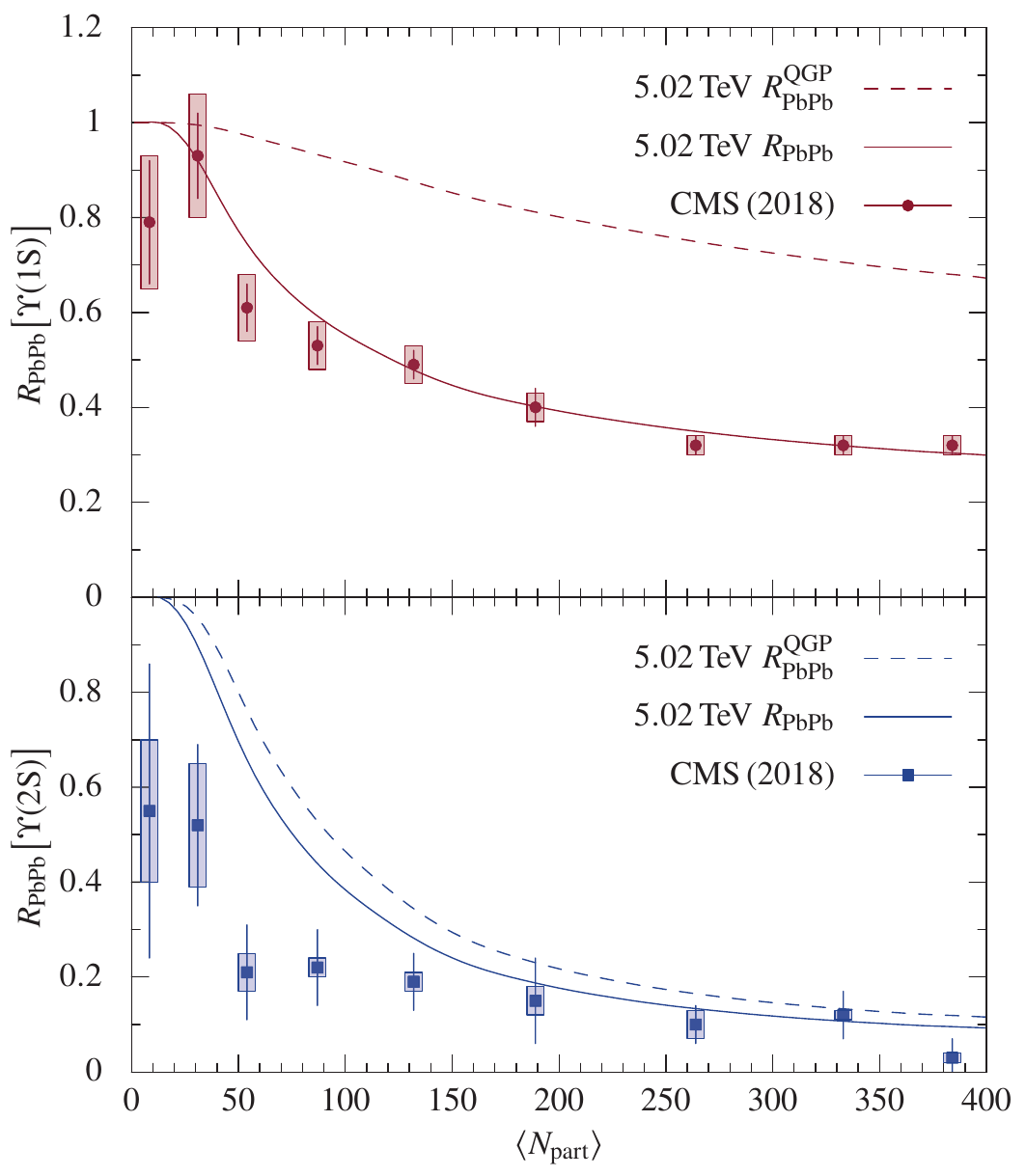}
\caption{(Color online) Top: Predicted modification factor $R_\text{PbPb}[\Upsilon(1S)]$ in Pb-Pb~collisions at $\sNN = {5.02}$ {TeV} (solid curve) with centrality-dependent data from CMS \cite{cms19} ($|y| < 2.4$) 
as function of the number of participants $\langle\Npart\rangle$ (averaged over centrality bins).
		The modification factor in the QGP-phase without the effect of reduced feed-down is shown as dashed (upper) curve. The
		formation time is $\tau_F=0.4$ fm/$c$, the initial central temperature $T_0=513$ MeV.
	Bottom: Predicted modification factor for the first excited state $R_\text{PbPb}[\Upsilon(2S)]$ in Pb-Pb~collisions at $\sNN = {5.02}$ {TeV}  (solid curve) with data from  CMS~\cite{cms19}. The modification factor in the QGP-phase (dashed) accounts for most of the calculated total suppression (solid) of the 
		$\YnS{2}$ state. Predictions are from Hoelck, Nendzig and Wolschin \cite{hnw17}.}
\label{fig11}
\end{figure}
The $\pT$-dependence of the $\YnS{1}$ and $\YnS{2}$ suppression as calculated in our model \cite{hnw17} is shown in Fig.\,\ref{fig8} 
for minimum-bias Pb-Pb~collisions at $\WithUnit{2.76}{TeV}$ together with CMS data~\cite{cms17}. 
The in-medium suppression factor $\RAAQGP$ for the ground state rises towards $\pT \simeq \WithUnit{10}{GeV/$c$}$ since it becomes easier for the bottomonia to leave the hot zone with increasing $\pT$.
At $\WithUnit{10}{GeV/$c$} \lesssim \pT \lesssim \WithUnit{20}{GeV/$c$}$ the rising widths overcompensate this trend, causing a fall of $\RAAQGP$.
When the reduced feed-down from the excited states is considered, the suppression factor $\RAA$ becomes rather flat, in reasonable agreement with the available CMS data for the ground state. For the $\Upsilon(1S)$ state, the contribution of reduced feed-down to the suppression is sizeable, whereas it is not significant for the $\Upsilon(2S)$ state \cite{hnw17}.

The calculation at 2.76 TeV is done for a formation time $\tau_\text{F} = 0.4$ fm/$c$, and we
have determined  the initial central temperature $T_0 = 480$ MeV from the observable under study.
This is in line with the historical expectation from $J/\psi$-physics that quarkonia dissociation
in the QGP should provide a thermometer to measure $T_0$, but it is, of course, a model-dependent result. 

Instead, it is meanwhile state of the art in the literature to use a bulk evolution calibrated on the light sector
and determine $T_0$ via flow measurements and their reproduction in 2D+1 viscous hydrocodes such as Ref.\,\cite{he16}, or even 3D+1 codes
that are available in ready-to-use packages before being adapted to heavy-flavour physics. However, these are also model-dependent results, and it would be
inconsistent to combine $T_0$-values from such codes with our approach that includes a less sophisticated 
background evolution.

The predicted energy dependence of the ground-state suppression at 5.02 TeV is also shown in Fig.\,\ref{fig8}: We found slightly ($\lesssim 10\%$)  more suppression 
compared to the 2.76 TeV result \cite{hnw17} when scaling the initial central temperature from $T_0=480$ MeV at the lower, to $T_0=513$ MeV at the higher energy.
Here, we have used the scaling relation between the initial entropy density and the charged particle multiplicity per unit rapidity, $s_0 \propto dN_\text{ch}/d\eta$ \cite{bjorken-1983,baym-etal-1983,gyulassi-matsui-1984}, and inserted $s_0 \propto T_0^3$ together with measured results \cite{alice17} for $dN_\text{ch}/d\eta$ to obtain an increase by 6.8\% in the initial temperature. 
In all calculations, the bottomonia formation time was kept fixed at $\tau_{\text{F},{nl}}=0.4$ fm/$c$ for all six states involved. The parameters for the density distribution of $^{208}$Pb are taken from Ref.\,\cite{vries87}.

The centrality dependence that we obtain \cite{hnw17} after averaging over $\pT$ is shown in Fig.\,\ref{fig9} for 2.76 TeV Pb-Pb, again together with a prediction at 5.02 TeV. The relative contributions of in-medium suppression and total suppression including reduced feed-down are exposed. At 2.76 TeV, $\simeq30$\% of the total $\YnS{1}$ suppression is due to the in-medium effects, but that fraction rises to $\simeq 40$\% at the higher energy. The situation is very different for the first excited state when calculated with the same set of parameters: The suppression in the QGP phase determines essentially the total suppression, reduced feed-down is unimportant \cite{hgw17}. This underlines the unique role of the $\YnS{1}$ state as a probe for bottomonia spectroscopy in heavy-ion collisions.

There is no rapidity dependence in our model for symmetric systems, both minimum-bias and centrality-dependent yields are flat as functions of $y$, corresponding to a boost-invariant hydrodynamical evolution.

Our predictions for the $p_\text{T}$-dependent $\Upsilon$ suppression in 5.02 TeV Pb-Pb collisions from Fig.\,\ref{fig8} are shown to be in agreement with CMS data \cite{cms19} in Fig.\,\ref{fig10}; see the caption for details. For the $\Upsilon\text{(1S)}$ state, a substantial fraction of the suppression, in particular at low $p_\text{T}$, is due to reduced feed-down as predicted. The corresponding centrality-dependent suppression (integrated over $p_\text{T}$) is shown in Fig.\,\ref{fig11}, in agreement with the data \cite{cms19} for the $\Upsilon\text{(1S)}$ state.
Related ALICE data at more forward rapidities $2.5<y<4$ are roughly consistent within the error bars \cite{alice19b}.
The prediction for the $\Upsilon\text{(2S)}$ modification factor underestimates the CMS data in peripheral collisions, which show strong suppression. We had found in Ref.\,\cite{hgw17} that the extra suppression is most likely not caused by the strong electromagnetic fields in more peripheral collisions, such that the origin remains presently unknown.

These results may be compared with those from some of the related approaches to $\Y$ suppression such as Refs.\,\cite{em12,striba12,peng11,song12}.
The model of Strickland and Bazow \cite{striba12} also includes dynamical propagation of the $\Y$~meson in the colored medium and a potential based on the heavy-quark internal energy and yields results that are consistent with the available data.
The strong binding model by Emerick, Zhao and Rapp \cite{em12} includes a contribution from cold nuclear matter effects and is also consistent with the data. 

The model by Song, Han and Ko \cite{song12} uses second-order gluon- and quark-dissociation of bottomonia rather than first order as here and 
        in other works such as \cite{peng11} Zhuang $et~al.$ They calculate in-medium production and dissociation 
from a rate equation. Wave functions and
decay widths are obtained from a screened Cornell potential that corresponds essentially to the real
part of the complex potential that we are using.
The fireball is modeled as a viscous, cylindrically symmetric fluid and
transversely averaged quantities are calculated. The inclusion of viscosity
allows for lower temperatures at the same QGP lifetime as compared to
perfect-fluid hydrodynamics in our modeling.
The two effects of bottomonium regeneration and gluonic  (anti-)shadowing
are also included in the model, but are found to have only little
impact on the results. The model by Ko $et~al.$ does, however, not include an imaginary part in the potential
to account for the significant contribution of collisional damping to the total width, and the running of the
strong coupling $\alpha_s$ is not considered.

Some of the various model results have been reviewed in comparison with data in Ref.\,\cite{an16} and Ref.\,\cite{qm19}. Once the respective parameters are tuned, the results are often found to be compatible with the data in spite of vastly different model ingredients (such as different quark-antiquark potentials) and hence, it is difficult to extract model-independent conclusions. Regarding the relative importance of in-medium suppression and feed-down for ground and excited states as function of energy that I have emphasized here, our conclusions should, however, be quite stable, and it would be interesting to test this proposition in the other models.
\section{Bottomonia in asymmetric systems}
\begin{figure}
\centering
\includegraphics[scale=0.26]{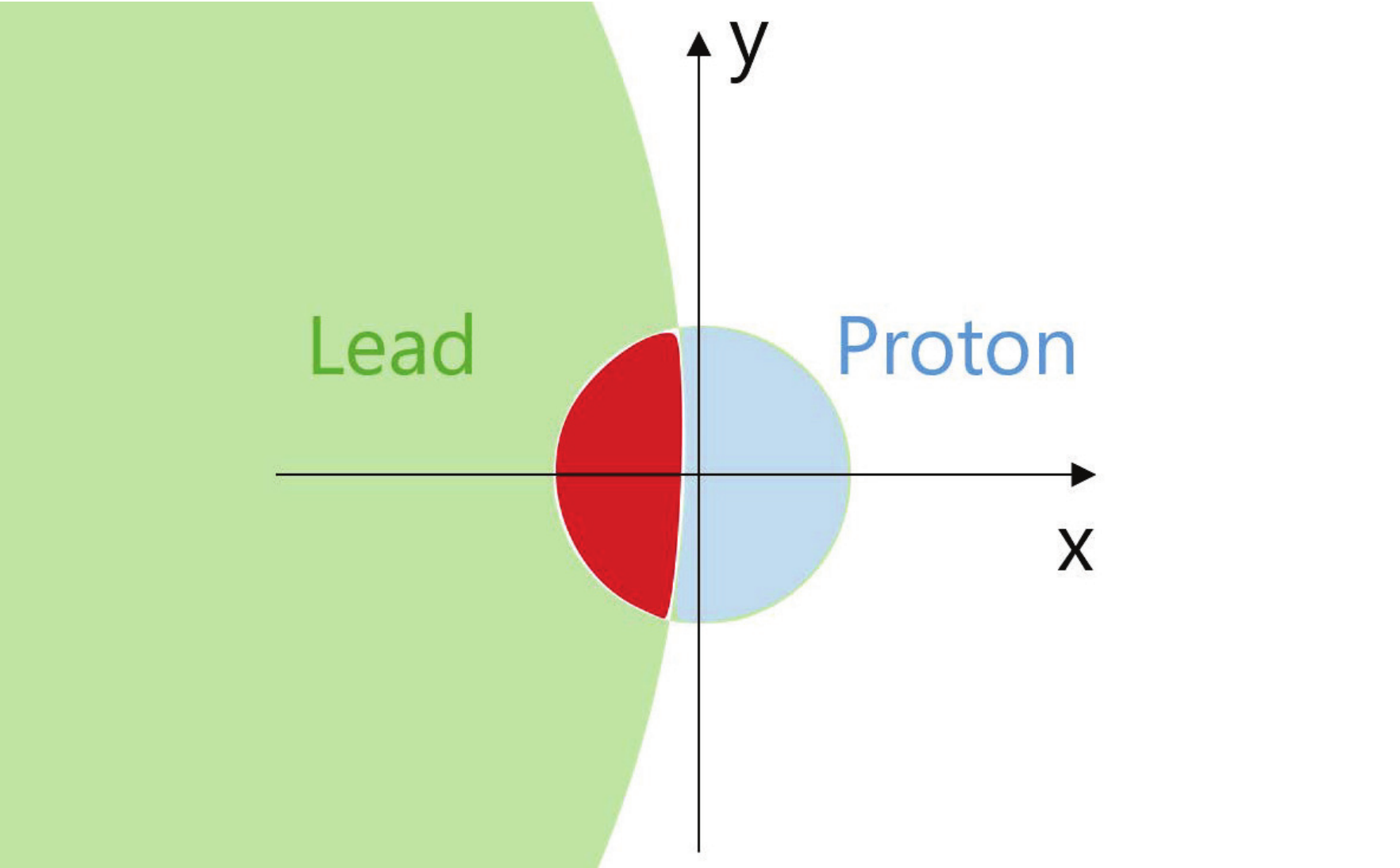}
	\caption{(Color online) Overlap (red) of the thickness functions in the transverse plane for lead (green) and proton (blue) in a peripheral collision. (From Dinh, MSc thesis HD 2019, unpublished. Reproduced with permission.)  }
\label{fig14}
\end{figure}

The asymmetric p-Pb system at $\sqrt{s_\text{NN}}$ = 8.16 TeV has been investigated experimentally by the LHCb\cite{lhcb18} and ALICE\cite{alice20} collaborations with respect to bottomonia production. 
The c.m. energy is achieved with a proton beam of the present maximum proton beam momentum of $p_\text{p}=6.5$ TeV according to
$\sqrt{s_\text{NN}}=2p_\text{p}\sqrt{Z_1 Z_2/(A_1 A_2)}$; with $p_\text{p}=4$ TeV, the c.m. energy per nucleon-nucleon pair is 5.02 TeV. The size relation and overlap function of projectile and target is indicated schematically in Fig.\,\ref{fig14}, here for a peripheral collision.
Although data at the lower energy are available, we confine the present comparison with our model to p-Pb collisions at the higher energy. 

\begin{figure*}
\centering
\includegraphics[scale=0.6]{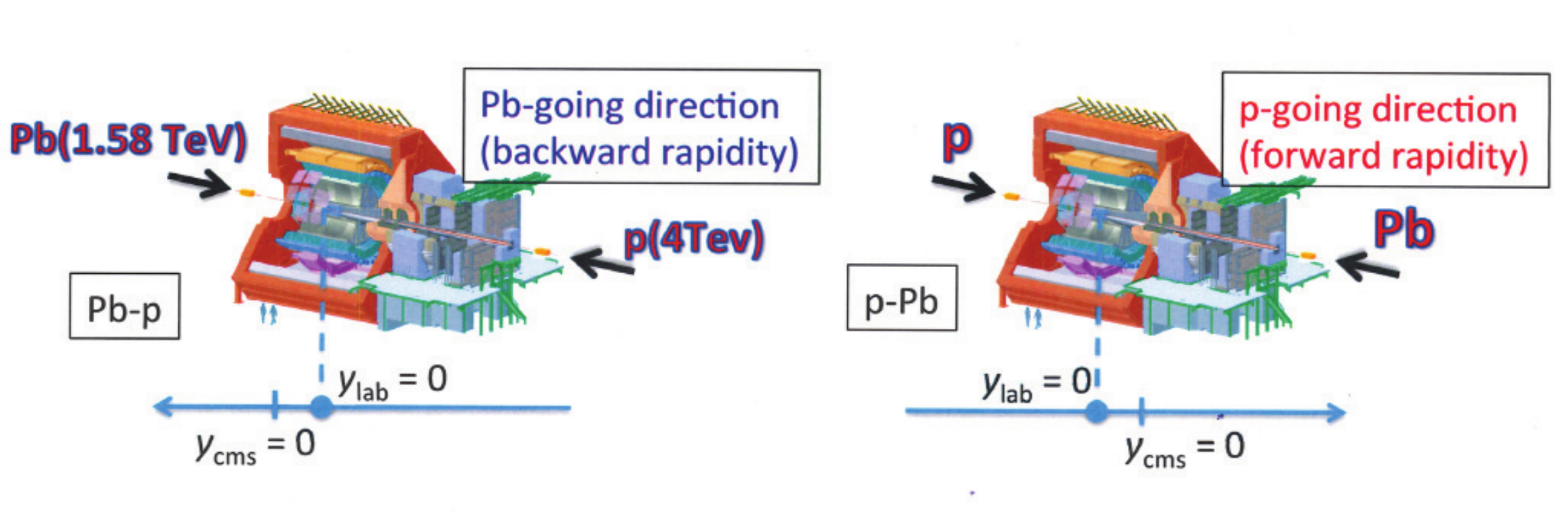}
	\caption{(Color online) Detector configuration for asymmetric systems and rapidity shift between laboratory and center-of-momentum systems. The indicated beam energies correspond to $\sqrt{s_\text{NN}}=5.023$ TeV. (\copyright ALICE Collaboration. Reproduced with permission by Dainese.)  }
\label{fig13}
\end{figure*}

To be able to measure both the forward (positive rapidities, p-going direction) and the backward (Pb-going) direction, the beams have to be interchanged as displayed in Fig.\,\ref{fig13}, and the rapidity shift $\Delta y=0.5\ln{(Z_1A_2/(Z_2A_1))}$ of the laboratory system relative to the center of momentum has to be considered when comparing data with calculations in the c.m. system. For p-Pb collisions, the shift is $\Delta y=0.465$, for d-Au $\Delta y=0.092$. 

\begin{figure}
\centering
\includegraphics[scale=0.6]{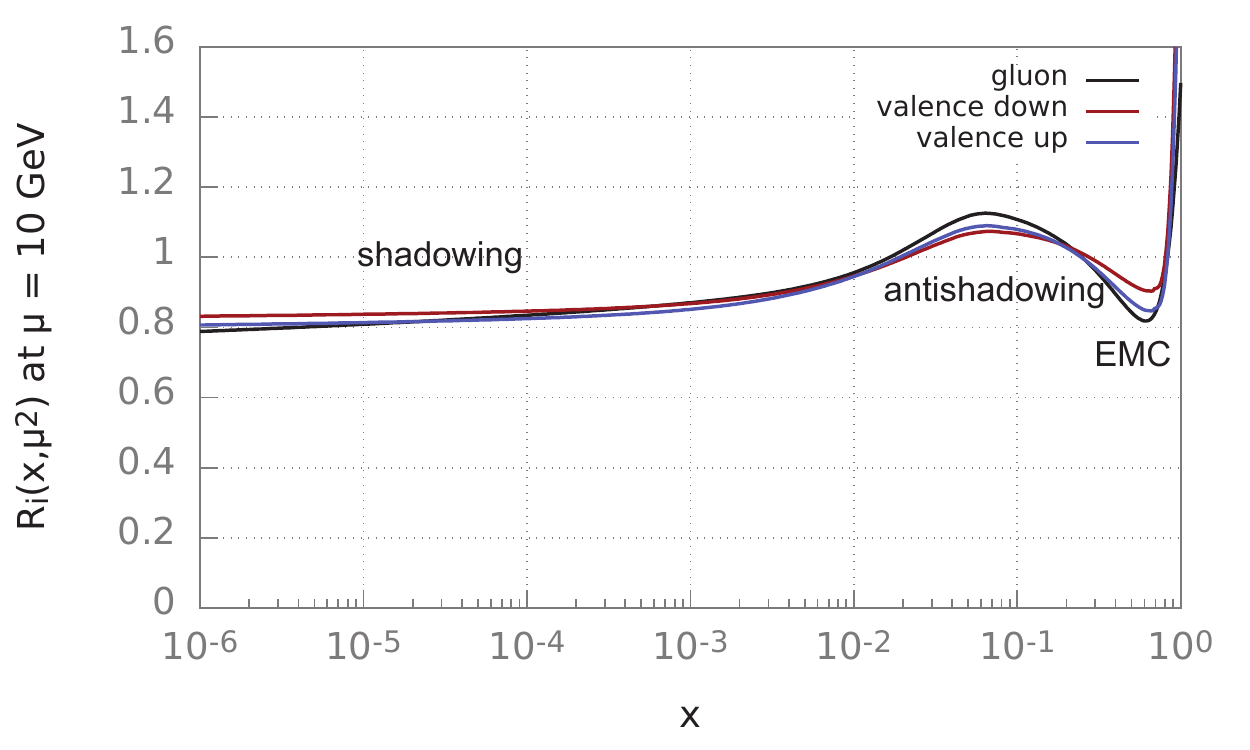}
	\caption{(Color online) Modification of the nuclear PDFs EPPS16 \cite{es17} for gluons, up- and down-quarks as function of the momentum fraction $x$ at $10^{-6} < x  \le 1$: Shadowing at $x \le 0.02$, antishadowing at $0.02<x<0.3$. (From Dinh, MSc thesis HD 2019, unpublished. Reproduced with permission.)  }
\label{fig15}
\end{figure}

Theoretically, cold nuclear matter effects in this system were studied in great detail in Ref.\,\cite{alba18}. 
These are much more relevant than in symmetric systems, because the bulk of the hadronic matter remains cold during the interaction. 
The dominant CNM effect is the modification of the parton distribution functions in the nuclear medium, which has been studied by many authors. A typical result for the PDF modifications with shadowing at small values of Bjorken-$x$, and antishadowing at intermediate $x$-values as obtained with \cite{es17}  EPPS16  is shown in Fig.\,\ref{fig15}. Shadowing causes a reduction of the $\Upsilon(nS)$ yields in p-Pb as compared to scaled pp, whereas antishadowing results in an enhancement. Shadowing is somewhat more pronounced if one, in addition, considers coherent energy-loss mechanisms in the cold medium. Still, these are not sufficient to interpret the available data in terms of CNM effects, as becomes obvious from direct comparisons, in particular, for the $\Upsilon(2S)$ state.

There is, however, a spatially small hot zone (fireball) with an initial central temperature that is comparable to the one in a symmetric system, and during its expansion and cooling, it contributes to bottomonia dissociation in regions where the temperature remains above the critical value. We have investigated the respective cold-matter and hot-medium effects on 
$\Upsilon$-dissociation in 8.16 TeV p-Pb collisions in Ref.\,\cite{dhw19}, treating the hot-medium dissociation with our model that has been originally developed for symmetric systems. In the following, some of our results are summarized. First the initial bottomonia populations and their modification in p-Pb as compared to pp collisions as well as coherent energy loss are discussed, then the hot-medium bottomonia dissociation is considered, and the results are compared with LHCb and ALICE data.
\subsection{Initial bottomonia populations in pp}
\label{sec:pops}

As indicated previously, the production of $\Y$ mesons in proton-proton collisions can occur either directly in parton scattering or via feed-down from the decay of heavier bottomonium states, such as $\chib$, or higher-mass $\Y$ states, thus complicating the theoretical description of bottomonium production.
In our model for p-Pb collisions, we make use of the measured double-differential pp cross sections of dimuon pairs at $\sqrt{s}=8$ TeV from $\Y$ decays, $\mathrm{d}^2\sigma_\text{pp}^{\Y\to\mathrm{\mu^+\mu^-}}\!/(\mathrm{d}p_\text{T}\mathrm{d}y)$, as obtained by the LHCb collaboration \cite{lhcb15}.
These data are rescaled using the corresponding dimuon branching ratios to obtain the inclusive bottomonium-decay cross sections in pp collisions, $\mathrm{d}^2\sigma_\text{pp}^{\Y\to X}\!/(\mathrm{d}p_\text{T}\mathrm{d}y)$.
Then, we apply an inverse feed-down cascade \cite{ngw13,vaccaro-etal-2013,hnw17} for every $p_\text{T}$ and $y$ bin to reconstruct the (direct) bottomonium-production cross sections in pp collisions, $\mathrm{d}^2\sigma_\text{pp}^\Y/(\mathrm{d}p_\text{T}\mathrm{d}y)$.
The latter do not include the indirect contributions from feed-down and are thus always smaller than the measured decay cross sections.
We fit decay and production cross sections separately with an analytical fit function proposed in Ref.\,\cite{arl13},
\begin{eqnarray}
	\frac{\mathrm{d}^2\sigma_\text{pp}}{\mathrm{d}p_\text{T}\mathrm{d}y}
	= \mathcal{N} \, p_\text{T} \left(\frac{p_0^2}{p_0^2 + p_\text{T}^2}\right)^{\!m} \left(1 - \frac{2M_\text{T}}{\sqrt{s}} \cosh y\right)^{\!n}.
	\label{eq:fitfunction}
\end{eqnarray}

\begin{figure}
\centering
\includegraphics[scale=0.76]{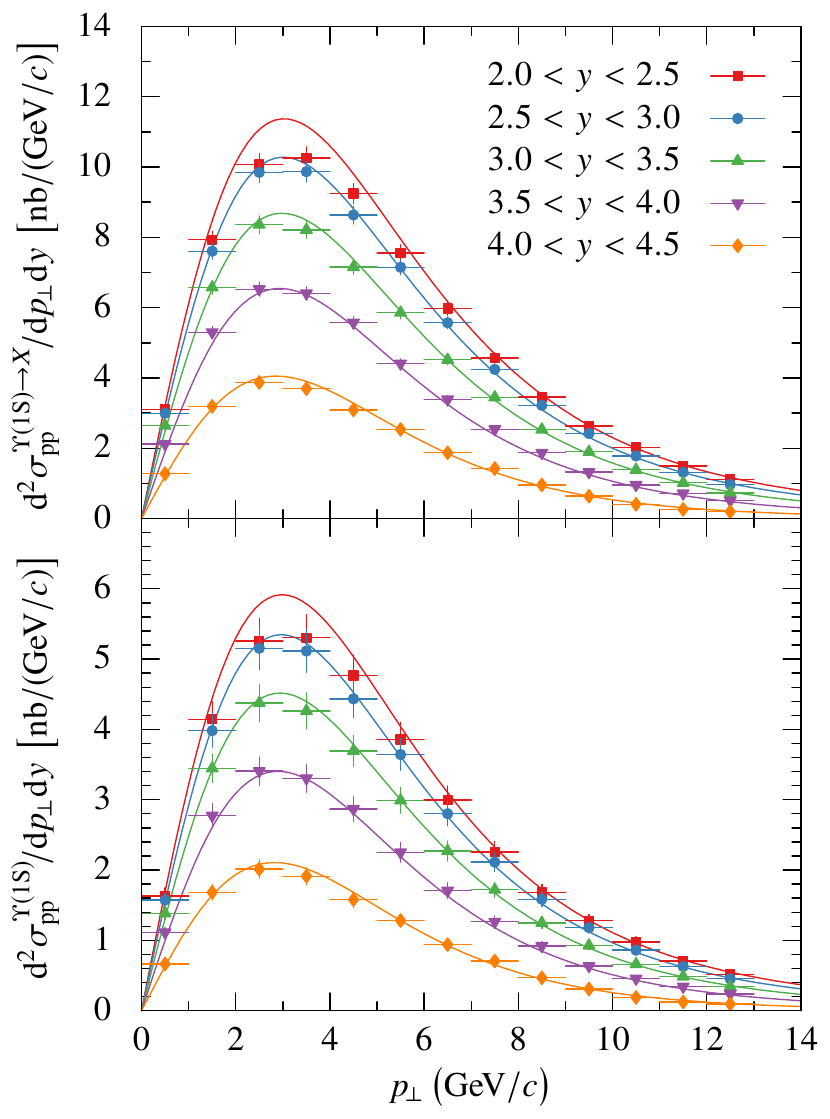}
	\caption{(Color online) 
		Fits of the double-differential cross sections for $\YnS{1}$ decays (top) and $\YnS{1}$ production (bottom) in $\sqrt{s}=8$ TeV pp collisions.
		The data points and error bars are based on LHCb data \cite{lhcb15}.
		Our corresponding fits are displayed as functions of transverse momentum for five rapidity regions. From Ref.\,\cite{dhw19}.}
\label{fig16}
\end{figure}
The fits are shown in Fig.\,\ref{fig16}.
Apart from the smallest rapidity bin, the fits are sufficiently precise, with an overall $\chi^2/\mathrm{ndf} = 2.12$ and $\chi^2/\mathrm{ndf} = 0.50$ for the $\YnS{1}$-decay and $\YnS{1}$-production cross sections, respectively.
Hence, we build the subsequent calculations for the $\YnS{1}$ and, similarly, $\YnS{2}$ yields in p-Pb collisions on the analytical functions.
\subsection{Cold-matter effects in asymmetric systems}
\label{sec:cnm}
In asymmetric collisions such as p-Pb, the bottomonia yields are already affected by the presence of nuclear matter -- even if a hot medium was completely absent.
These well-known cold nuclear matter effects include initial-state effects such as the modification of the initial gluon densities, as well as mixed initial- and final-state effects such as the coherent parton energy loss induced by the nuclear medium.
We have considered in Ref.\,\cite{dhw19} the above two CNM-effects and determined the corresponding modification of bottomonia yields in p-Pb as compared to what is expected from pp collisions at LHC energies.

The modification from pp to p-Pb collisions is quantified by the nuclear modification factor
\begin{equation}
	R_\text{pPb}(b, p_\text{T}, y) = \frac{1}{\Ncoll(b)}\frac{\frac{\mathrm{d}^2\sigma_\text{pPb}^{\Y\to X}}{\mathrm{d}p_\text{T}\mathrm{d}y}(b,p_\text{T},y)}{\frac{\mathrm{d}^2\sigma_\text{pp}^{\Y\to X}}{\mathrm{d}p_\text{T}\mathrm{d}y}(p_\text{T},y)}~,
\end{equation}
where $\mathrm{d}^2\sigma^{\Y\to X}\!/(\mathrm{d}p_\text{T}\mathrm{d}y)$ is the Lorentz-invariant double-differential cross section for $\Y$~decays.
For p-Pb, these cross sections are calculated via the decay cascade from the corresponding production cross sections after applying all cold-matter and hot-medium modifications.
$R_\text{pPb}$ depends on the rapidity~$y$, transverse momentum~$p_\text{T}$, and the centrality of the collision which can be expressed either in terms of the impact parameter~$b$ or by the average number of binary nucleon-nucleon collisions~$\Ncoll(b)$, which is 15.6 for a central p-Pb collision \cite{dhw19}, and drops monotonically towards peripheral collisions.  In minimum-bias collisions, the value is $\Ncoll(b=7\,\text{fm})=2.8$.
Modification factors as functions of only one observable are obtained by integrating the differential cross sections first before taking the ratio,
\begin{eqnarray}
	\label{eq:pTFactor}
	R_\text{pPb}(p_\text{T}) &= \frac{\iint\frac{\mathrm{d}^2\sigma_\text{pPb}^{\Y\to X}}{\mathrm{d}p_\text{T}\mathrm{d}y}\frac{\mathrm{d}\sigma_\text{pPb}^\text{inel}}{\mathrm{d}b}\mathrm{d}b\,\mathrm{d}y}{\int\Ncoll\frac{\mathrm{d}\sigma_\text{pPb}^\text{inel}}{\mathrm{d}b}\mathrm{d}b \, \int\frac{\mathrm{d}^2\sigma_\text{pp}^{\Y\to X}}{\mathrm{d}p_\text{T}\mathrm{d}y}\mathrm{d}y}\,,
\end{eqnarray}
and $R_\text{pPb}(y)$ accordingly.
Here, ${\mathrm{d}\sigma_\text{pPb}^\text{inel}}/{\mathrm{d}b}$ is the differential inelastic p-Pb cross section which is given by the Glauber model \cite{bialas1976}.

For our calculations, we consider the nuclear gluon shadowing and the coherent parton energy loss model to account for the CNM effects.
Thus, the CNM modification of the bottomonium-production cross section from pp to p-Pb collisions reads \cite{arl13,vogt15,dhw19}
\begin{eqnarray}
	\label{eq:AllEffects}
	\frac{1}{\Ncoll} \frac{\mathrm{d}^2\sigma_\text{pPb}^\text{CNM}}{\mathrm{d}p_\text{T} \mathrm{d}y}=\qquad\qquad\qquad\qquad\qquad\qquad\qquad\\\nonumber 
	 \int_0^{2 \pi} \frac{\mathrm{d}\varphi}{2 \pi} \int_0^{\varepsilon_\text{max}} \mathrm{d}\varepsilon \,
	P(\varepsilon, E, L_\text{eff}) \,
	\frac{p_\parallel}{p_\parallel^\text{shift}}\frac{p_\text{T}}{p_\text{T}^\text{shift}} \,
	R_g^\text{Pb}(x_2^\text{shift})\\  \nonumber
	\times\frac{\mathrm{d}^2\sigma_\text{pp}}{\mathrm{d}p_\text{T} \mathrm{d}y}(p_\text{T}^\text{shift}, y^\text{shift})
\end{eqnarray}
where the shifted quantities are expressed by
\begin{eqnarray}
	\label{eq:RapShift}
	y^\text{shift} &=
	 \operatorname{arcosh}\!\left[\frac{E(p_\text{T},y) + \varepsilon}{M_\text{T}(p_\text{T}^\text{shift})}\right] - y_\text{beam}\,,\\
	p_\text{T}^\text{shift} &= \sqrt{{p_\text{T}^2 + \small{\Delta}p_\text{T}^2 + 2 p_\text{T} \small{\Delta}p_\text{T} \cos \varphi}}\,,\\
	p_\parallel^\text{shift} &= \sqrt{{\big[E(p_\text{T},y) + \varepsilon\big]^2 - M_\text{T}^2(p_\text{T}^\text{shift})} }\,.
	\label{eq:MomShift}
\end{eqnarray}
The integral is over the energy loss $\varepsilon$ and the angle $\varphi$ between the transverse momentum of the bottomonium and the total transverse momentum kick $\small{\Delta}p_\text{T}$ in the lead nucleus.
It is based on the model of Arleo, Peign\'{e} $et~al.$ for parton energy loss \cite{arl13,arl16}, where
partons traversing a medium are expected to loose energy via induced gluon radiation caused by interactions with multiple static scattering centers of the medium.

In the above expression, the number of produced bottomonia in p-Pb collisions is thus calculated from the production of higher energetic bottomonia in pp collisions and the probability that they will emit the energy difference $\varepsilon$ through gluon radiation.
The probability distribution for the energy loss $\varepsilon$ of bottomonia with energy $E$ in the lead rest frame is given by \cite{arl13} the normalized quenching weight $P(\varepsilon, E, L_\text{eff})$.
The transverse-momentum kick $\small{\Delta}p_\text{T}$ is related to the gluon saturation scale in the lead nuclei, which is of the order of 1 GeV/$c$. The shifted variables in Eqs.\,(\ref{eq:RapShift}) - (\ref{eq:MomShift}) follow from kinematic considerations.

We have included the gluonic nuclear modification factor $R_g^\text{Pb}$ of the gluon PDF in Pb (see Fig.\,\ref{fig15}) in Eq.\,(\ref{eq:AllEffects}).
This is a simplification of the shadowing effects in the Color Evaporation Model as formulated by Vogt \cite{vogt15}, since we assume that the main contribution to the bottomonium-production cross section comes from gluon fusion.
Hence, the bottomonium momentum fraction $x_2$ is given by the kinematics of $2 \rightarrow 1$ processes,
\begin{eqnarray}
	x_2(p_\text{T}, y) &= \frac{M_{\Y,\text{T}}}{\sqrtsNN}\exp(-y) \,, \\
	x_2^\text{shift} &= x_2(p_\text{T}^\text{shift}, y^\text{shift}) \,,
\end{eqnarray}
where $M_{\Y,\text{T}}$ is the transverse mass of the final-state bottomonium.
We have used the EPPS16 set \cite{es17} which includes a global analysis of nuclear shadowing.

Since we have considered both, shadowing and coherent energy loss in the cold nuclear matter, we must adapt the value of the transport coefficient $\hat{q}$ that governs the energy-loss model of Arleo $et~al.$ \cite{arl13,arl16}, reducing it from $\hat{q} =$ 0.075 to 0.046 GeV$^2$/(fm\,$c^2$).
The corresponding difference in the nuclear modification factors is, however, small because the transverse-momentum kick $\Delta p_\text{T}$ scales with $\sqrt{\hat{q}}$\,:
The modification is below 3\% in $p_\text{T}$-averaged results and becomes significant only at $p_\text{T} < 5$ GeV/$c$.

The centrality dependence of the CNM modification factor is caused by the changing effective path length $L_\text{eff}$ in  Eq.\,(\ref{eq:AllEffects}), which in turn affects the quenching weight $P(\varepsilon, E, L_\text{eff})$ in 
Eq.\,(\ref{eq:AllEffects}).
The path length for a projectile travelling through a medium is proportional to the number of binary collisions and the mean free path.
The latter is given by the inverse of the product of the inelastic pp cross section $\sigma_\text{pp}^\text{inel}$ and the mean number density $\rho_0$ in the nucleus,
\begin{eqnarray}
	\label{eq:EffectivePathLength}
	L_\text{eff}(b) = \frac{\Ncoll(b)}{\rho_0 \ \sigma_\text{pp}^\text{inel}}\,,
\end{eqnarray}
with the mean number density $\rho_0\approx 0.17\ \text{fm}^{-3}$.
Using the results from our Glauber calculation and \cite{ant13} $\sigma_\text{pp}^\text{inel}(8\, \text{TeV}) \simeq 7.46$ fm$^2$, we obtain the value
\begin{eqnarray}
	L_\text{eff, Pb}(0) \approx 12.26\,\text{fm}
\end{eqnarray}
in central collisions, which scales with $\Ncoll(b)$ for more peripheral collisions.

Our calculations for the CNM effects due to both modifications of the PDFs and coherent energy loss in the nuclear medium \cite{dhw19} are in line with standard results of the CNM-community \cite{alba18}, which were originally presented as predictions before the 8.16 TeV p-Pb run. In the next section, we proceed to investigate the influence of the quark-gluon droplet in asymmetric systems at LHC energies on the $\Y$ suppression.

\section{Hot-medium effects in p-Pb collisions}
\label{sec:qgp}
The initial QGP-zone in p-Pb collisions at LHC energies -- here, at $\sqrtsNN = 8.16$ TeV-- is spatially much less extended compared to symmetric systems like Pb-Pb. It has turned out, however, that the dissociation of bottomonia states in the hot medium is significant and can not be neglected.
In Ref.\,\cite{dhw19} we have adapted our model for hot-medium bottomonia suppression in symmetric collisions to the case of asymmetric systems.
Again, the bottomonia states are taken to be produced with a formation time \cite{ngw14,hnw17} $\tauF \simeq 0.4$ fm/$c$  in initial hard collisions at finite transverse momentum $p_\text{T}$, and then move in the hot expanding medium made of gluons and light quarks where the dissociation processes take place. A possible temperature  and state dependence of the formation time has been discussed in Sec.\,\ref{ingredients}.

The local equilibration time of the QGP droplet is very short, about 0.1 fm/$c$ for gluons \cite{fmr18} and less than 1 fm/$c$ for quarks, such that the conditions for a hydrodynamic treatment of the expansion and cooling of the hot zone are fulfilled, as in a symmetric system.
The difference in the local equilibration time for quarks versus gluons has been discussed on nonequilibrium-statistical grounds in Ref.\,\cite{gw18}.
It is essentially due to the role of Pauli's principle, but the different color factors will enhance it. 
\begin{figure}
\centering
\includegraphics[scale=0.64]{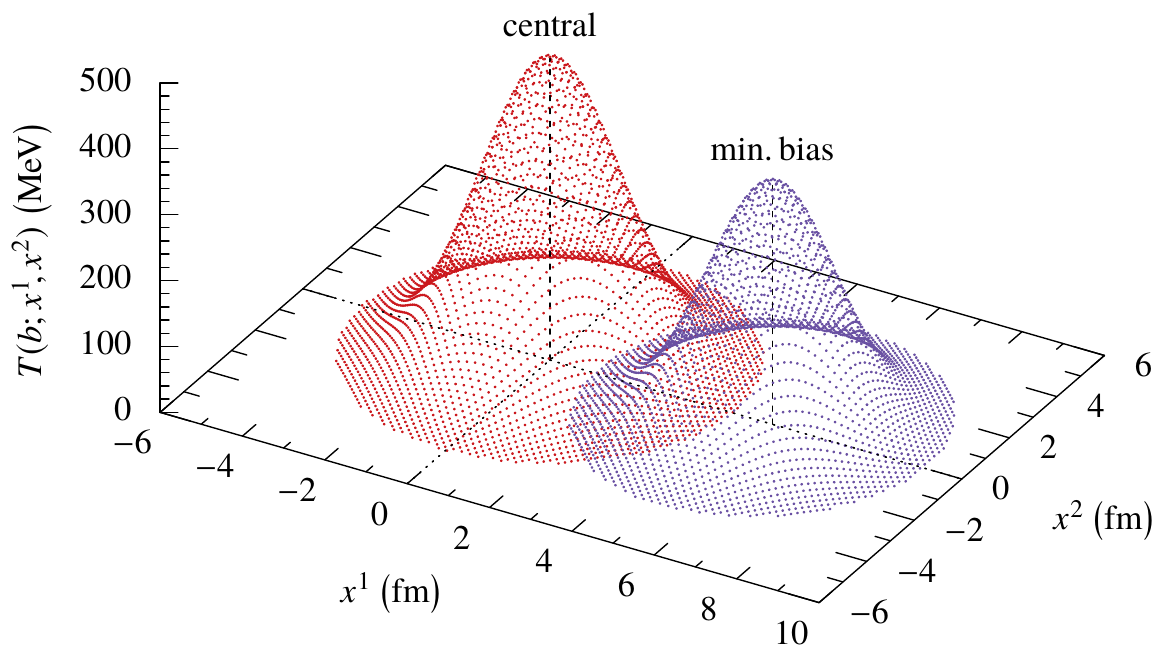}
	\caption{(Color online) Initial temperature profiles of the hot QGP generated in p-Pb collisions at $\sqrt{s_\text{NN}} = 8.16$ TeV as functions of the transverse coordinates $(x^1,x^2)$ at two centralities: Central collisions with $N_\text{coll} \simeq 15.6$, left, and minimum-bias collisions with $N_\text{coll} \simeq 7$, right.  The temperature exceeds the critical value in a sizeable region of the transverse plane. From Ref.\,\cite{dhw19}.}
\label{fig17}
\end{figure}

As in case of symmetric systems \cite{ngw14,hnw17}, we use perfect-fluid relativistic hydrodynamics with longitudinal and transverse expansion to account for the background bulk evolution. Following the outline in Sec.\,\ref{expansion}, we have solved \cite{dhw19}
the equations of motion
for the four-velocity~$u$ and the temperature distribution~$T$ in the transverse plane $(x^1,x^2)$ numerically, starting at the initial time 0.1 fm/$c$ in the LCF.
For asymmetric systems, we adapt the initial condition for $T$ to scale with the distribution of binary collisions in the transverse plane $\ncoll(b;x^1,x^2)$ that lead to the formation of the hot zone,
\begin{equation}
	T(b; \tau_{\text{init}}, x^1, x^2) = \Tinit \, \sqrt[3]{\frac{\ncoll(b;x^1,x^2)}{\ncoll(0;0,0)}}\,.
\end{equation}
In p-Pb collisions, the expected number of binary collisions in a central collision is $\Ncoll(b{\,=\,}0) \simeq 15.6$, where $\Ncoll(b) = \int\mathrm{d}^2x \, \ncoll(b;x^1,x^2)$.

As discussed in the next section where some of our comparisons with data will be shown, we have determined the initial central temperature~$\Tinit$ in p-Pb collisions at 8.16 TeV by fitting our cold-matter plus hot-medium results to LHCb data at forward rapidities, resulting in $\Tinit=460$ MeV. It will be interesting to see if future independent determinations result in compatible values.
For example, it may be feasible to determine the initial central temperature from a comparison of hydrodynamic calculations with experimental results for elliptic flow of charged hadrons \cite{ALICE-2019}.
It would, however, be less reliable as in case of large symmetric systems, where flow is a more pronounced property.
The inclusion of viscosity would alter our results slightly, allowing for lower temperatures at the same QGP lifetime as compared to perfect-fluid hydrodynamics in our modeling.

The distribution of binary collisions~$\ncoll(b;x^1,x^2)$ has been obtained from a Glauber calculation and is proportional to the nuclear overlap function~$\theta_\text{pPb}$,
\begin{gather}
	\theta_\text{pPb}(b;x^1,x^2) = \theta_\text{p}(b;x^1,x^2) \times \theta_\text{Pb}(b;x^1,x^2)\,,\\[2ex]
	\theta_\text{p}(b;x^1,x^2) = \int\mathrm{d}x^3 \rho_\text{p}(|b\vec{e}_1 - \vec{x}|)\,,\\
	\theta_\text{Pb}(b;x^1,x^2) = \int\mathrm{d}x^3 \rho_\text{Pb}(|\vec{x}|)\,,
\end{gather}
where $\rho_\text{p},\rho_\text{Pb}$ are the radial symmetric nucleon distributions of the proton and lead nucleus, respectively.
 We have taken a Woods-Saxon potential with parameters from Ref.\,\cite{vries87} for the latter and a Gaussian shape for the proton with a corresponding radius of 0.875 fm.
\begin{figure*}
	\centering
  \centering
  \includegraphics[scale=0.46]{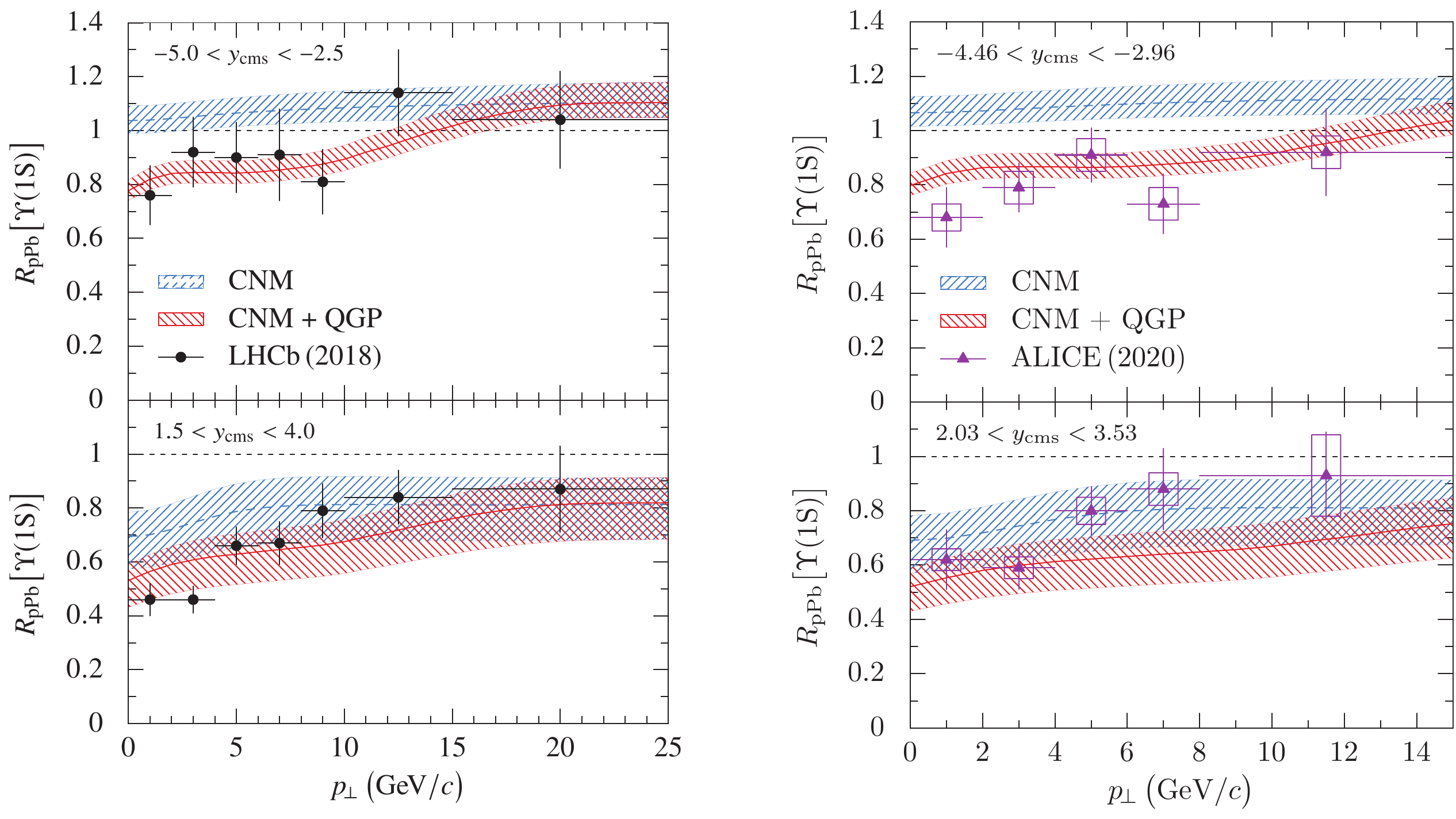}
  \caption{(Color online) Left panel: Calculated $p_\text{T}$-dependent nuclear modification factors $R_\text{pPb}$ for the $\YnS{1}$~spin-triplet ground state in p-Pb~collisions at $\sqrt{s_\text{NN}} = 8.16$ TeV  are compared with LHCb data \cite{lhcb18} in the backward (Pb-going, top) and forward (p-going, bottom) region, for minimum-bias centrality.
		Results for CNM effects that include shadowing, energy loss, and reduced feed-down (dashed curves, blue) are shown together with calculations that incorporate also QGP effects (solid curves, red). Right panel: Our calculations \cite{dhw19} are compared with ALICE data \cite{alice20} in a slightly different rapidity range, and for $p_\text{T} < 15$ GeV.
		The error bands result from the uncertainties of the parton distribution functions that enter the calculations.}
  \label{fig18a}
  \end{figure*}
\begin{figure*}
	\centering
	\includegraphics[scale=0.46]{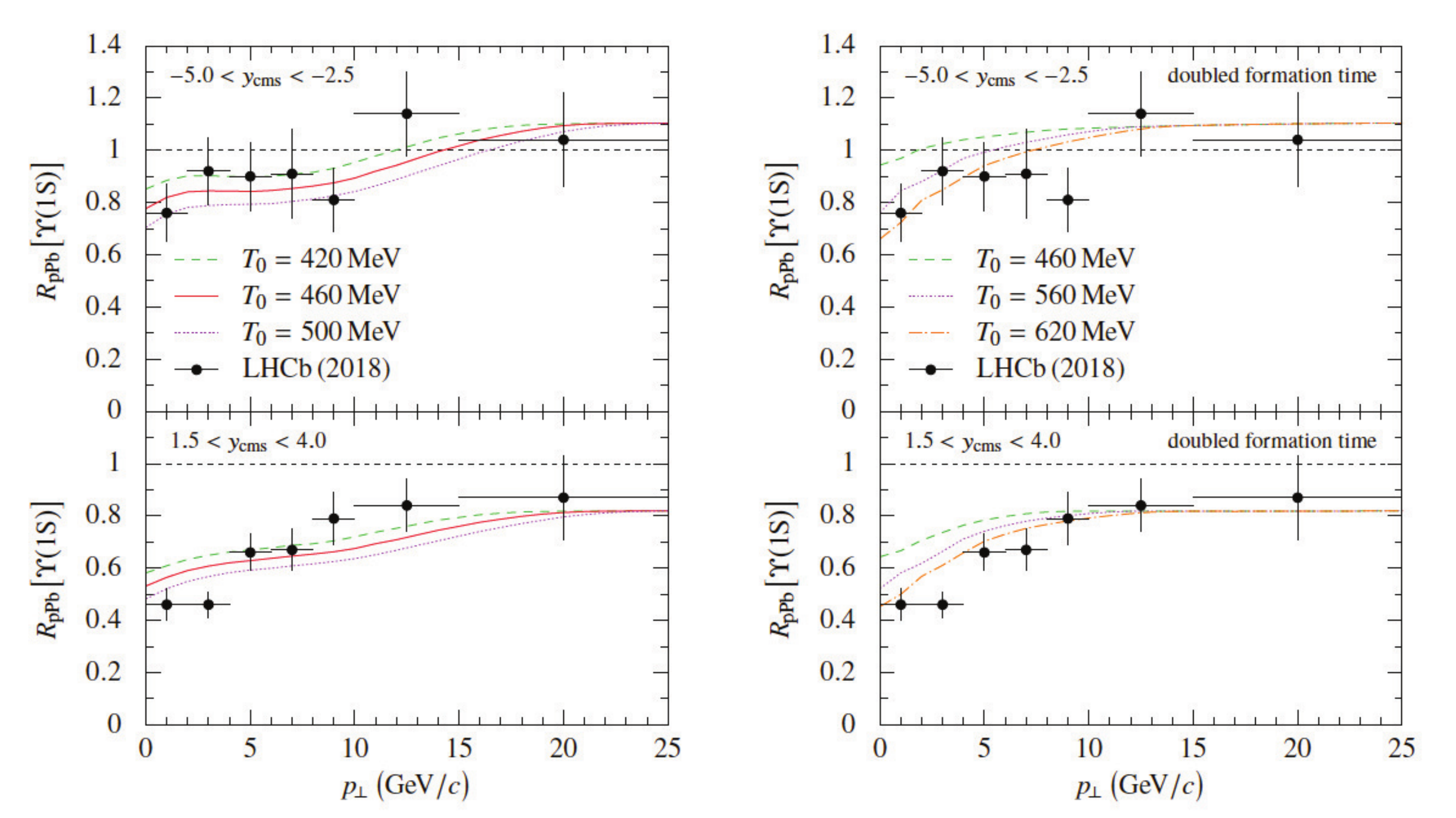}
	\caption{
		(Color online)
		Left panel: Dependence of the nuclear modification factors $R_\text{pPb}$ for the $\YnS{1}$ state in p-Pb~collisions at  $\sqrt{s_\text{NN}} = 8.16$ TeV  on the initial central temperature $T_0$ in the backward (Pb-going, top) and forward (p-going, bottom) region, with formation time $\tau_\text{F} = 0.04$ fm/$c$.
		CNM and QGP effects are included. Right panel: Results for $R_\text{pPb}[\YnS{1}]$ in p-Pb~collisions at  $\sqrt{s_\text{NN}} = 8.16$ TeV  when doubling the bottomonia formation time to $\tau_\text{F} = 0.08$ fm/$c$, for minimum-bias centrality and three values of the initial central temperature $T_0$.
		Data are from Ref.\,\cite{lhcb18}, the calculations from Ref.\,\cite{dhw19}.}
		\label{fig18x}
\end{figure*}
\begin{figure}
  \centering
  \includegraphics[scale=0.8]{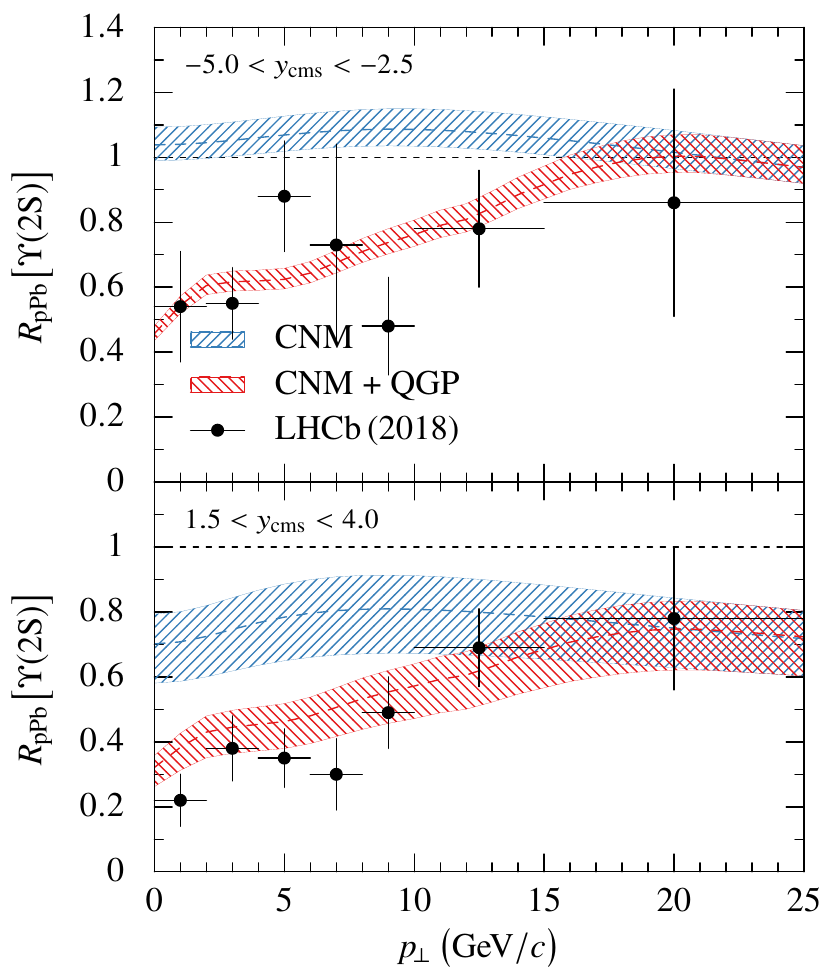}
    \caption{(Color online) Calculated $p_\text{T}$-dependent nuclear modification factors \cite{dhw19} $R_\text{pPb}$ for the $\YnS{2}$~spin-triplet first excited state in p-Pb~collisions at $\sqrt{s_\text{NN}}=8.16$ TeV are compared with LHCb data \cite{lhcb18} in the backward (Pb-going, top) and forward (p-going, bottom) region, for minimum-bias centrality.
		Results for CNM effects that include shadowing, energy loss, and reduced feed-down (dashed curves, blue) are shown together with calculations that incorporate also QGP effects (solid curves, red).
		The error bands result from the uncertainties of the parton distribution functions that enter the calculations. From Ref.\,\cite{dhw19}.}
  \label{fig19}
\end{figure}
Two-dimensional initial temperature profiles in the transverse $(x^1,x^2)$-plane for p-Pb with the above parameters are shown in Fig.\,\ref{fig17} for two values of the impact parameter corresponding to central ($b=0$) and minimum-bias ($\Ncoll(b) = \operatorname{MinBias}\left[\Ncoll\right] \simeq 7$) collisions.
Although the hot zone with $T > \Tcrit \simeq 160$ MeV is substantially less extended in p-Pb as compared to Pb-Pb, it is still sufficiently pronounced to cause in-medium dissociation of the initially produced bottomonia states.

To obtain the hot-medium decay widths of the relevant bottomonia states, the energies $E_{nl}(T)$ and corresponding damping widths $\Gamma_{\text{damp},{nl}}(T)$ as a function of QGP temperature~$T$ are calculated as described in Sec.\,\ref{ingredients} for symmetric systems, starting from a solution of Schr\"odinger's equation for the bottomonia at each point in the transverse plane 
with a complex, temperature-dependent potential $V_{nl}(r,T)$  for the six states $\YnS{n}$ and $\chibnP{n}$, $n=1,2,3$, using an iterative method to account for the running of the strong coupling. In addition, 
 the width caused by gluon-induced dissociation \cite{bgw12,ngw14}  $\Gamma_{\text{diss},{nl}}(T)$ has been calculated as detailed in Sec.\,\ref{gluodissociation}.
 It is added incoherently to the damping width.

As discussed already in case of symmetric systems, the bottomonia are not expected to be co-moving with the expanding hot medium in the transverse plane due to the high bottom-quark mass. Consequently, there is a 
 finite relative velocity between medium and bottomonia, and we have applied
 the relativistic Doppler effect to the medium temperature in the bottomonium rest frame and performed an angular average over the shifted decay widths as explained in Ref.\, \cite{hnw17}.

\section{Comparison to data for asymmetric systems}
\subsection{Transverse-momentum dependence}

\begin{figure}
	\begin{center}
	\includegraphics[scale=0.82]{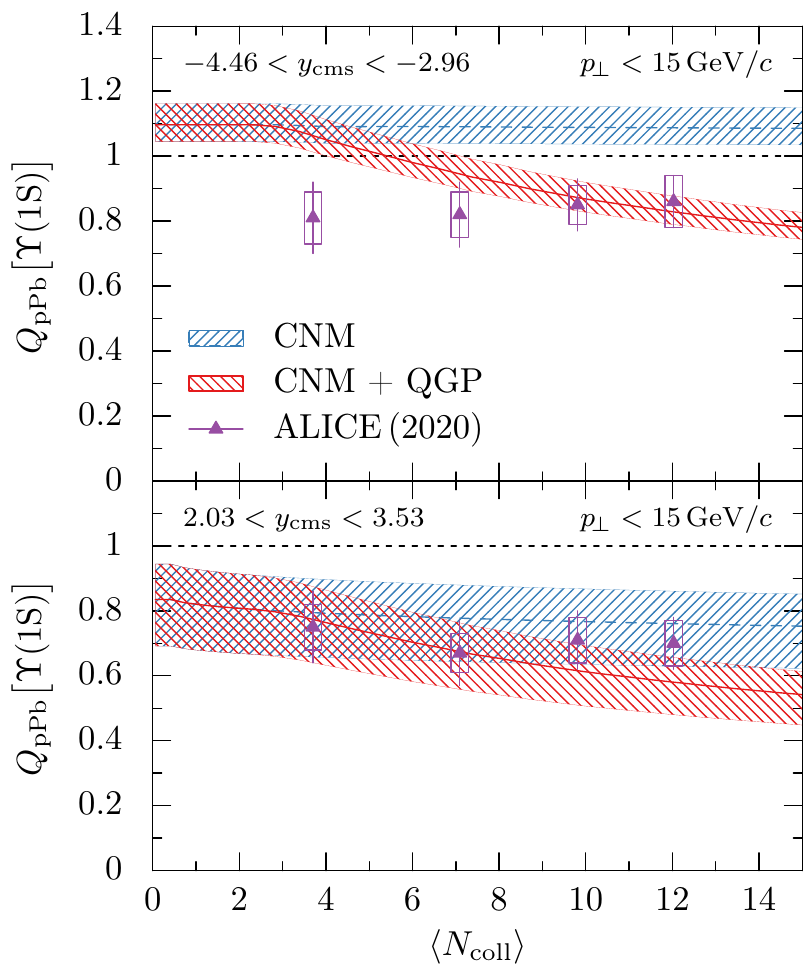}
	\caption{
		(Color online)
		Calculated centrality-dependent nuclear modification factors \cite{dhw19} $R_\text{pPb}$ for the $\YnS{1}$~state in p-Pb~collisions at $\sqrtsNN = 8.16$ TeV in the backward (Pb-going, top) and forward (p-going, bottom) region are compared with  ALICE data denoted as \cite{alice20} $Q_\text{pPb}$ (see text).
		The rapidity regions are as in Fig.\,\ref{fig18a}.
		Results for CNM effects that include shadowing, energy loss, and reduced feed-down (dashed curves, blue) are shown together with calculations that incorporate also QGP-droplet effects (solid curves, red).
		The error bands result from the uncertainties of the parton distribution functions that enter the calculations.}
		\label{fig19a}
		\end{center}
\end{figure}

\begin{figure}[tph]
\begin{center}
\includegraphics[scale=0.82]{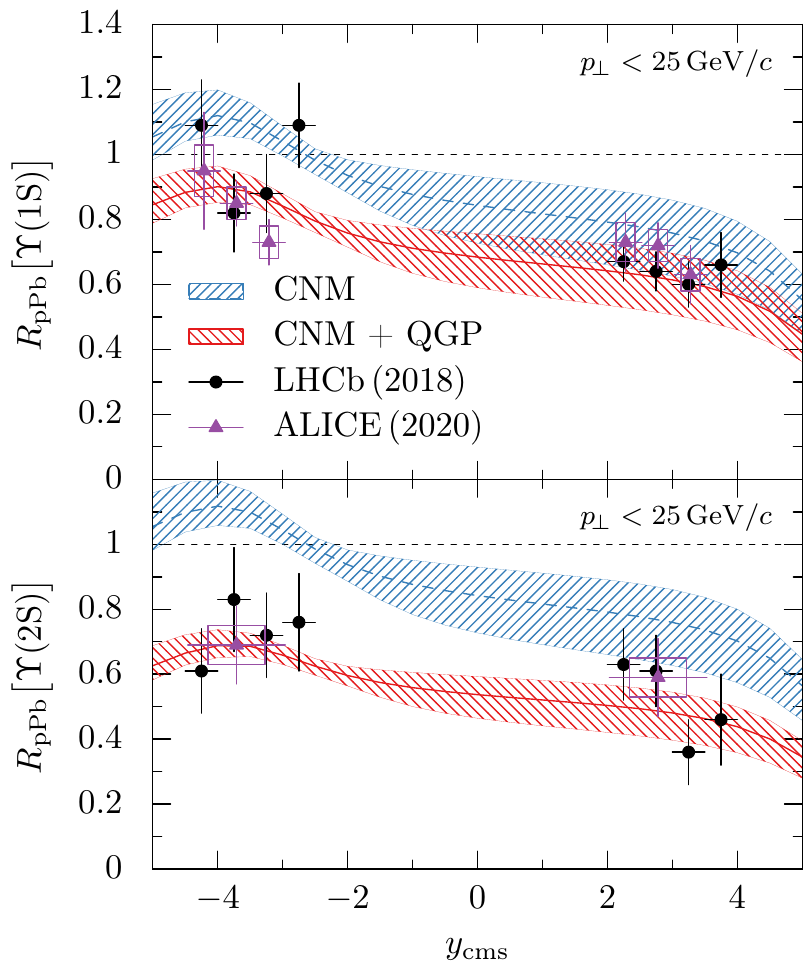}
	\caption{(Color online) Calculated rapidity-dependent nuclear modification factors $R_\text{pPb}$ for the $\Upsilon\text{(1S)}$ (top) and $\Upsilon\text{(2S)}$~state (bottom) in p-Pb~collisions at 
	$\sNN = 8.16$ TeV with LHCb data, circles \cite{lhcb18}, and ALICE data, triangles \cite{alice20}. For the latter, the maximum transverse momentum is 15 GeV/$c$.
		Results for cold nuclear matter (CNM) effects that include shadowing, coherent energy loss, and feed-down (dashed curves, blue) are shown together with calculations that incorporate also QGP effects (solid curves, red).
		The error bands result from the uncertainties of the parton distribution functions that enter the calculations. The initial central temperature in the QGP droplet region is $T_0\simeq 460$ MeV. Calculations from Dinh, Hoelck and Wolschin \cite{dhw19}, with added $\Upsilon\text{(2S)}$ LHCb data \cite{lhcb18}, and updated ALICE data \cite{alice20}.}
		\label{fig20}
		\end{center}
\end{figure}
Representative results from this work are shown in
Fig.\,\ref{fig18a} for the transverse-momentum dependence of the  $\Upsilon(1S)$ and in Fig.\,\ref{fig19} of the $\Upsilon(2S)$ state, with the forward (p-going) region in the lower and the backward (Pb-going) region in the upper frame, respectively. The plots show CNM (blue, upper bands)  and CNM plus QGP (red, lower bands) effects on the $\Upsilon(1S)$ and $\Upsilon(2S)$ yields in 8.16 TeV p-Pb collisions at the LHC. The transverse-momentum dependence in the backward direction (top) indicates for both states enhancement
due to antishadowing when only the CNM effects are considered, whereas the data \cite{lhcb18} for $\Upsilon(1S)$ are clearly suppressed at $p_\text{T} <10$ GeV/$c$ and for $\Upsilon(2S)$ at all measured $p_\text{T}$ values. 

This severe discrepancy in the backward region is cured through the consideration of the momentum-dependent dissociation in the QGP as shown in our cold-matter plus hot-medium calculation (red) \cite{dhw19}. In the forward region, already the CNM calculations show suppression for both states, but the agreement with the data is improved once the hot-matter contribution is added.

The hot-medium effects are seen to be different forward and backward, because the hot zone in the transverse
plane is more extended in the Pb-going region, thus causing more suppression in the backward region.

The dependence of the modification factor on the initial central temperature in the range $420\le T_0 \le 500$ MeV is shown in Fig.\,\ref{fig18x} (left panel), with $T_0 = 460$ MeV as the best result. Because the temperature is proportional to the third root of the charged-hadron multiplicity, the dependence is rather weak. It would, of course, be useful to have an independent estimate from other observables, as is the case in symmetric systems, where the temperature can be inferred from the comparison of $v_2$-measurements with hydrodynamical calculations. We have also tested the dependence of the hot-medium effects on the bottomonium formation time in Ref.\,\cite{dhw19}, with the result (see right panel in Fig.\,\ref{fig18x}) that doubling the formation time yields about 20\% less suppression, because the $\Upsilon$'s are formed at a later stage when the fireball droplet has already cooled down.

\subsection{Centrality dependence}
The calculated centrality dependence of $R_\text{pPb}$  for the $\YnS{1}$ state is displayed in Fig.\,\ref{fig19a} together with the ALICE data \cite{alice20}. The latter are quantified as \cite{alice20} $Q_\text{pPb}$ to account for potential biases from the centrality estimation that are unrelated to nuclear effects. Regarding the calculation, CNM effects result in a fairly flat dependence on the number of binary collisions both backward -- where antishadowing enhances $R_\text{pPb}$ above one -- and forward, where shadowing and energy loss already cause suppression.
The hot-medium contributions generate even more suppression in central collisions.
This result for $R_\text{pPb}$  disagrees, however, with the ALICE data for $Q_\text{pPb}$ which show almost no centrality dependence in the forward region, and backward even a slight rise of $Q_\text{pPb}$ with increasing centrality.
The origin of the discrepancy is an open question, but it may be due to the experimental bias in the centrality estimate.
ALICE data for the $J/\psi$ modification factors in 5.02 TeV p-Pb collisions show an even stronger rise towards $Q_\text{pPb}\simeq 1.2$ with increasing centrality backward  \cite{adam15}, although there is growing suppression in the forward region.
\subsection{Rapidity dependence}
When displayed as function of rapidity (Fig.\,\ref{fig20}), the asymmetric shape of the nuclear modification factors for both, $\Upsilon(1S)$ and $\Upsilon(2S)$ states, arises mainly from the different cold-matter effects in the forward and backward regions,
in particular, shadowing of the parton distribution functions at $y>0$ and antishadowing at $y<0$, but also coherent energy loss in the relatively cold medium. The additional suppression due to the dissociation in the hot QGP droplet is shown in the lower (red) curves, which are in better agreement with the data for the $\Upsilon\text{(1S)}$ ground state not only in the backward, but also in the forward direction. The substantial role of the hot-medium effects is even more pronounced for the $\Upsilon\text{(2S)}$  first excited state, where the CNM-calculation shows enhancement in the backward region, whereas the full calculation with in-medium dissociation displays a suppression down to almost 70\%  -- in agreement with the LHCb data \cite{lhcb18} and the ALICE data point from Ref.\, \cite{alice20}. The consistency with the data in the forward region is also improved for $\Upsilon(1S)$ once the hot-medium effects are included. With CNM effects alone, it appears to be impossible to achieve a consistent interpretation of the asymmetric-system data.

There is, however, no unanimous consensus yet
of the community regarding my conclusion on the presence of a QGP state in p-Pb.
In the works of Ferreiro and Lansberg \cite{fela18}, the discrepancy between CNM calculations and data for the $\Upsilon(nS)$ suppression in p-Pb has been explained in terms of interactions with comoving hadrons, in particular pions. We have not included this process in our calculations -- initially on the grounds that interactions of the bottomonia states with comovers were found to be unimportant at LHC energies in the work of Ko et al.\,\cite{ko01} about $\Upsilon$ absorption in hadronic matter. Probably one eventually has to consider both comover interactions and suppression in the hot QGP zone in order to fully understand the $\Upsilon$ modification data in asymmetric systems. 

In asymmetric systems, the feed-down cascade from the excited bottominia states to the ground state produces some additional $\Upsilon(1S)$ suppression due to the depopulation of the excited states -- but by far not as much as in symmetric systems, where the excited states are almost totally screened or depopulated, thus substantially reducing the feed-down to the ground state.

\section{Conclusions}
Our phenomenological model for $\Upsilon$ suppression in relativistic heavy-ion collisions incorporates screening, damping,  gluodissociation, and reduced feed-down. With the parameters bottomonia formation time and initial central temperature fixed, we have used it to predict \cite{hnw17} the \us\, suppression together with its transverse-momentum and centrality dependence in Pb-Pb collisions at $\sNN = {5.02}$ {TeV} accurately when compared to recent CMS data \cite{cms19}.
Screening is found to be unimportant for the \us\ state, whereas reduced feed-down is responsible for a considerable fraction of the ground-state suppression. 

In contrast, the model reveals substantial screening effects for the excited \uss~state  and -- together with the other dissociation processes that we consider -- more suppression than for \us, with only a small contribution from reduced feed-down. In very peripheral collisions, however, the current CMS data \cite{cms19} for $\Upsilon\text{(2S)}$ show more suppression than the model, leaving room for future improvement. We had shown that electromagnetic field effects \cite{hgw17} are unlikely to be the origin of the discrepancy because the lifetime of the transient fields is too short, even when the time prolongation through the medium is considered.

Regarding bottomonia in asymmetric systems, we have investigated p-Pb collisions at the LHC center-of-momentum energy of 8.16 TeV per particle pair \cite{dhw19}. For an adequate understanding of the data, we need to consider not only the well-established CNM effects, but also the hot-medium suppression in the spatially small, but rapidly expanding, fireball. This result is consistent with the presence of a quark-gluon droplet in asymmetric systems at LHC energies, as proposed in Ref\,\cite{phenix19} for RHIC energies. It is, however, at present not yet reflecting the unanimous consensus of the community in this field.

\begin{acknowledgments}
	The original works underlying this brief review were performed in collaboration with the Heidelberg PhD students Johannes H\"olck (ITP Heidelberg), Felix Nendzig (now at Accso-Accelerated Solutions GmbH, Darmstadt, Germany), and Viet Hung Dinh (now IJC Paris-Saclay) who contributed in his MSc thesis CNM results for modifications of $\Upsilon$ yields in p-Pb collisions at LHC energies. I am grateful to K. K. Phua for inviting me to write this article.\end{acknowledgments}
\bibliography{gw_20}

\end{document}